\documentclass[apjl]{emulateapj}

\usepackage{graphics}      
\usepackage{amsmath}
\usepackage{amssymb}
\usepackage{morefloats}

\def\sles{\lower2pt\hbox{$\buildrel {\scriptstyle <}
   \over {\scriptstyle\sim}$}}
\def\sgreat{\lower2pt\hbox{$\buildrel {\scriptstyle >}
   \over {\scriptstyle\sim}$}}

\begin{document}
 
\title{THEORETICAL TRANSIT SPECTRA FOR GJ 1214b AND OTHER “SUPER-EARTHS”}

\author{Alex R. Howe and Adam S. Burrows\altaffilmark{1}} 

\altaffiltext{1}{Department of Astrophysical Sciences, 
Peyton Hall, Princeton University, Princeton, NJ 08544, USA; arhowe@astro.princeton.edu, burrows@astro.princeton.edu}

\begin{abstract}

We present new calculations of transit spectra of super-Earths that allow for atmospheres with arbitrary proportions of common molecular species and haze.  We test this method with generic spectra, reproducing the expected systematics and absorption features, then apply it to the nearby super-Earth GJ 1214b, which has produced conflicting observational data, leaving the questions of a hydrogen-rich versus hydrogen-poor atmosphere and the water content of the atmosphere ambiguous.  We present representative transit spectra for a range of classes of atmosphere models for GJ 1214b.  Our analysis supports a hydrogen-rich atmosphere with a cloud or haze layer, although a hydrogen-poor model with $\lesssim$10\% water is not ruled out.  Several classes of models are ruled out, however, including hydrogen-rich atmospheres with no haze, hydrogen-rich atmospheres with a haze of $\sim$0.01 ${\rm \mu m}$ tholin particles, and hydrogen-poor atmospheres with major sources of absorption other than water.  We propose an observational test to distinguish hydrogen-rich from hydrogen-poor atmospheres.  Finally, we provide a library of theoretical transit spectra for super-Earths with a broad range of parameters to facilitate future comparison with anticipated data.

\end{abstract}


\section{Introduction}
\label{intro}

Ground- and space-based transit surveys are discovering a growing number of exoplanets in the super-Earth range, defined alternately as between 1 and 10 Earth masses or between 1 and 2 Earth radii.  The largest such survey, being conducted by the {\it Kepler} spacecraft, has already announced hundreds of super-Earth candidates (Kepler Objects of Interest) and one confirmed super-Earth in the ``habitable zone'' \citep{BoruckiII}, although only a few of these candidates have been confirmed.  Nevertheless, these surveys have yielded several exoplanets for which it is possible to begin characterizing their atmospheres via transit spectroscopy.  This has been attempted by several groups for GJ 1214b \citep{Bean2011,Berta II,Croll,Crossfield,Desert,de Mooij}, an estimated 6.55 Earth-mass ($M_\Earth$) planet orbiting an M-dwarf star with an equilibrium temperature near 500 K \citep{Charbonneau}, with conflicting results.  In particular, measured transit depths in the $K$ and $K_s$ bands are in direct conflict, and proposed atmospheric models have been unable to fit observations in other bands simultaneously.

GJ 1214's faintness ($m_V = 14.67$) makes it a challenge for spectroscopic observations.  However, the recent discovery of transits of 55 Cnc e around a naked-eye star (Winn et al. 2011) holds promise for improved spectroscopic characterizations in the near future.  These stars bring their own challenges of smaller transit depths and fewer nearby comparisons stars, but efforts are nonetheless underway to characterize them.

While transit and radial velocity observations can together constrain a planet's mass and radius (and thus its density), this leaves significant degeneracies in its bulk composition.  Planets with similar masses and radii can take on different compositions, ranging from a mini-Neptune with a thick H/He layer to a true super-Earth composed mostly of silicates with a high-molecular-weight atmosphere to a ``water world'' whose bulk composition and atmosphere both have high water content \citep{Valencia,Fortney,Seager,Sotin,Rogers}.  To further confuse matters, recent observations suggest that some giant exoplanets may have significantly higher C/O ratios than their parent stars, opening up the additional possibility of carbon-rich compositions \citep{Madhusudhan}.

GJ 1214b faces the additional complication that while its measured mass and density suggest a hydrogen-dominated atmosphere, the measured radius of the parent star, GJ 1214, is significantly larger than predicted by stellar models \citep{Charbonneau}.  If the star is 15\% smaller than was measured, as these stellar models predict, GJ 1214b would be correspondingly smaller and denser (given the same transit depth), and a hydrogen-rich atmosphere would not be indicated.  This uncertainty means that we cannot definitively distinguish between a hydrogen-rich and a hydrogen-poor atmosphere without spectroscopic data \citep{Miller-Ricci2010,Carter}.

Spectroscopic observations of both secondary eclipses (``emission'') and transits (``transmission'') of exoplanets have proven successful in characterizing the atmospheres of gas giants, e.g. \citep{Burrows2009,Charbonneau2002,Deming,Knutson,Swain}, and observations to characterize super-Earths continue.  Transit spectra are particularly useful for measuring compositions.  While secondary eclipses can be used to measure temperatures and compositions, the transit spectrum is most directly sensitive to the column depth at the limb of individual molecular and atomic species and is only weakly dependent on temperature (for a given scale height).  In this paper, we present calculations of transit spectra for GJ 1214b for a variety of atmospheres in order to improve the current fits to the data.  We also provide a library of models for future exploration of transit spectra of super-Earths.

Transit spectroscopy measures the apparent size of a planet in various wavelengths, which depends on the averaged properties of the planet's atmosphere.  Because exoplanets cannot be resolved by current telescope technology, this transit spectrum is a measure of the total amount of starlight blocked in transit at each wavelength.  Ignoring the effects of limb darkening, this includes a fixed component due to the solid disk of the planet, plus a wavelength-dependent component absorbed or scattered by the atmosphere.  Within the atmosphere, different altitudes will contribute different amounts to this component because of the variations with pressure.  However, molecular absorption features should still emerge from these spectra in a predictable way, enabling abundance characterization of planetary atmospheres based on transit observations.  Conversely, the absence of these features in the observed spectrum might be indicative not only of their absence, but of a cloud or haze layer, as is inferred in the atmosphere of HD 189733b \citep{Pont}, but one should keep an open mind that a small scale height that reduces the cross section of the atmosphere could also cause this effect.  The molecular features can be characterized by the relative increase in transit depth when compared with the continuum (or with a local minimum in transit depth if the continuum is obscured), a contrast that we hereafter refer to as the ``amplitude'' of the spectral features.

If the atmosphere is non-isothermal or not vertically mixed, this further complicates the results.  Also, the limb of the planet as seen in transit represents the terminator, where there is likely to be circulation and a change in temperature and composition.  Indeed, the physics at the terminator is more complicated than any model to date.  However, one-dimensional theoretical spectra are not strongly dependent on these factors and provide valuable zeroth-order approximations to estimate atmospheric compositions \citep{Miller-Ricci2009}.  In our models, hydrogen-rich atmospheres are taken to be in chemical equilibrium, while hydrogen-poor atmospheres are taken to be vertically mixed.

The physical parameters of an exoplanet should have predictable effects on its transit spectrum.  These mainly act through the scale height of the atmosphere, which is given by
\begin{equation}
H = \frac{kT}{\mu g},
\end{equation}
where $k$ is Boltzmann's constant, $T$ is the atmospheric temperature, $\mu$ is the mean molecular weight, and $g$ is the surface gravity.  If the atmosphere is approximately isothermal, $T$ is the ``equilibrium temperature.''

The amount of starlight absorbed in transit is proportional to the depth of the atmosphere and, thus, to its scale height.  This should be visible in the transit spectra as proportional changes in the amplitudes of the spectral features and the slope of the Rayleigh continuum.  Therefore, these amplitudes should be directly proportional to the temperature of the atmosphere and inversely proportional to its mean molecular weight.  To first order, the amplitudes of spectral features should also be inversely proportional to surface gravity, but this relation is not exact since gravity changes slightly with altitude.  In very thick atmospheres, the lower gravity at high altitudes should make the scale height larger, resulting in an even larger atmosphere and larger amplitudes than a simple estimate might suggest.

The remaining parameters that affect the transit spectrum are related to the composition structure of the atmosphere.  In particular, for a hydrogen-rich atmosphere, metallicity is important.  While for a hydrogen-rich atmosphere metallicity does not have a strong effect on mean molecular weight, the relatively large absorption opacities of metal compounds compared with molecular hydrogen mean that significantly larger-amplitude spectral features will be found at larger metallicities than at smaller ones.  The addition of clouds or haze should have the opposite effect.  The wavelength dependence of extinction opacity for haze is usually smaller than for molecular species, so the corresponding spectral features should be weaker.  In the simplest case of a uniformly opaque cloud layer, suggested by Croll et al. (2011) and Bean et al. (2011) for GJ 1214b, only molecular absorption above the cloud tops contributes to the variation in the spectrum, and the molecular features are suppressed.  All of these relations are seen in our calculated spectra.

We present new calculations of theoretical transit spectra of extrasolar super-Earths, including Mie scattering and a range of molecular compositions.  Modeling of transit spectroscopy has been established for super-Earths previously, for example, in Miller-Ricci \& Fortney (2010), and it has been used to try to characterize the atmosphere of GJ 1214b.  However, we present a number of new results and models including spectra for sub-solar metallicities, enhanced C-to-O ratios, and explicit calculations for a range of haze models.  We also outline a library of models for super-Earths with different parameters.  We present our methodology in Section \ref{method}.  We present general conclusions and demonstrations of the systematics described above in Section \ref{general}.  We apply our method to GJ 1214b in Section \ref{gj1214} in an attempt to produce a better fit to observations than current models.  We then present some general conclusions in Section \ref{conclusion}.

\section{Methodology}
\label{method}

For each atmosphere model, we produce a theoretical spectrum by computing the transit depth at 2500 frequency points between 0.3 and 10 ${\rm \mu m}$.  At each frequency point, we numerically integrate the optical depth along the line of sight of each impact parameter, then integrate the effective area over all impact parameters.  For transiting planets, we use the initial observed transit depth as a fiducial value for the radius and then normalize the spectrum by computing the average transit depth over the observational bandpass, equating this to the observed depth.  Without independent knowledge of the radius of the solid surface, a procedure like this is necessary for all models and modelers.

We use an isothermal temperature-pressure ({\it T-P}) profile at the equilibrium temperature.  However, we set $T = 470$ K for GJ 1214b based on the work of Miller-Ricci \& Fortney (2010), who find with more ambitious calculations that this is approximately correct over the pressure range probed by transit spectroscopy, although their approach does not fully capture the conditions at the terminator and should not be considered more reliable than any other model.  Additionally, transit spectra have been shown not to be strongly dependent on the exact T-P profile used \citep{Miller-Ricci2009}.  We do account for the variation of gravity with altitude when computing the pressure profile, although this effect is only a few percent in transit depth, depending on the scale height of the atmosphere.  We implement haze using the Mie approximation for a monodispersed, constant abundance haze between upper and lower pressure levels.  We also include the option of completely opaque cloud decks by cutting off transmission at all frequencies at pressures greater than a specified level \citep{Crossfield}.

\begin{figure}[htp]			
\begin{center}
\includegraphics[scale=0.9]{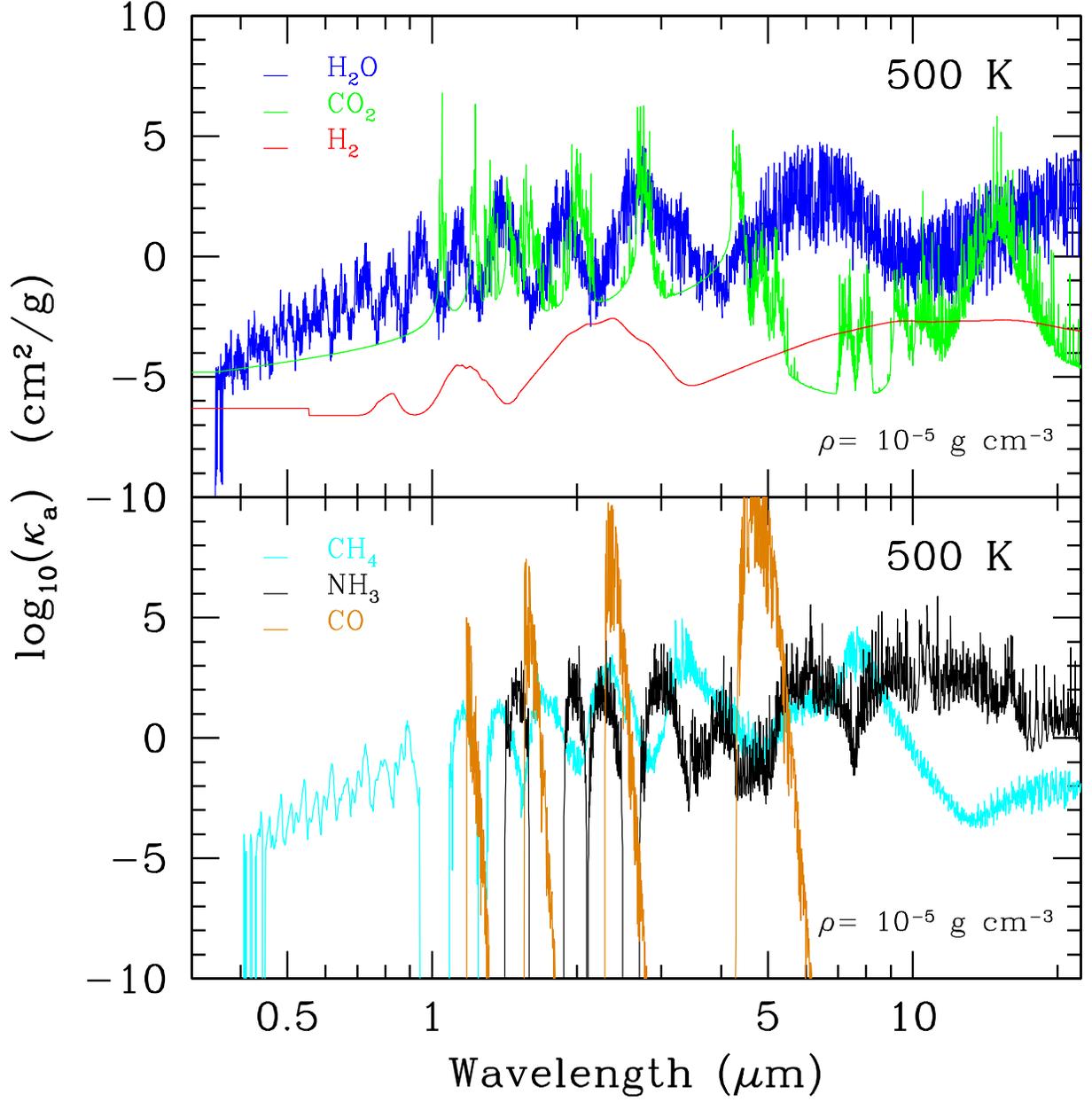} 
\end{center}
\caption{Absorptive opacities of the pure molecular species ${\rm H_2O, \, CO_2, \, H_2, \, CH_4, \, NH_3, \, and \, CO}$ in ${\rm cm^2 \, g^{-1}}$ plotted against wavelength in microns at a temperature of 500 K and a total atmospheric density of $10^{-5} \, \mathrm{g \, cm^{-3}}$, corresponding to a pressure of ~1 mbar, which is the approximate pressure probed by transit spectroscopy.  Vertical discontinuities extending past the bottom of the plot represent wavelengths at which the species in question is assumed to have zero absorption opacity.  Scattering opacity is not included.}
\label{opacities}
\end{figure}

Our chemical equilibrium abundances for molecular species were taken from Burrows \& Sharp (1999) and molecular line and collisional opacities were taken from Sharp \& Burrows (2007).  We compute Rayleigh cross sections on a per-molecule basis, based on each species’ electric polarizability and assuming a simple $\lambda^{-4}$ power-law dependence.  For haze components, we use Mie scattering theory for several possible haze species.  Polyacetylenes are polymers based on CH units produced by UV photolysis of simple hydrocarbons, the simplest being ${\rm (CH)_n}$, and are suggested to occur on Jupiter and giant exoplanets \citep{Sudarsky}.  Tholins are more complex heteropolymers produced by UV photolysis of hydrocarbons, nitrogen, and oxygen and are thought to occur in the atmosphere of Titan and on the surfaces of other outer solar system bodies \citep{Khare}.  They take the form of cross-linked chains of a wide variety of organic subunits, including aromatic hydrocarbons, other rings, alkanes, and polypeptides, with no definite order \citep{Sagan}.  We also consider sulfuric acid because of its presence on Venus.  Complex indices of refraction were taken from Khare et al. (1984) for tholins, from Bar-Nun et al. (1988) for polyacetylenes, and from Palmer \& Williams (1975) for sulfuric acid.  Our absorption opacities for pure molecular species at a temperature and density representative of GJ 1214b are shown in Figure \ref{opacities}.  The extinction opacities of hazes of various sizes and compositions are shown in Figure \ref{haze-xsec}.

We calculate transit spectra for a variety of atmospheric compositions.  The hydrogen-rich compositions we compute include metal abundances varying from 0.01 to 30 times solar, solar abundances with methane set to zero, and otherwise-solar abundances with the abundances of carbon and oxygen reversed, resulting in a C/O ratio of $\sim$2.  We consider an atmosphere with no methane because methane is easily destroyed by UV photolysis, and it has been suggested as a possibility to explain the appearance of GJ 1214b's spectrum \citep{Miller-Ricci2010,Bean2011,Desert}.  We consider an elevated C/O ratio because such a possibility has been advanced for giant planets, including those with a low C/O ratio in the host star \citep{Madhusudhan} and for terrestrial planets \citep{Kuchner}.  The C/O ratio is especially important because it strongly affects the abundances of common molecular species such as ${\rm H_2O, CO, CO_2, and \, CH_4}$.

For hydrogen-poor models, we use vertically mixed atmospheres that can include $\mathrm{H_2}$, $\mathrm{CO_2}$, $\mathrm{H_2O}$, $\mathrm{CH_4}$, $\mathrm{NH_3}$, $\mathrm{CO}$, and  $\mathrm{N_2}$ in arbitrary proportions by volume.  We assume nitrogen to have Rayleigh scattering opacity only.

\begin{figure}[htp]			
\begin{center}
\includegraphics[scale=0.9]{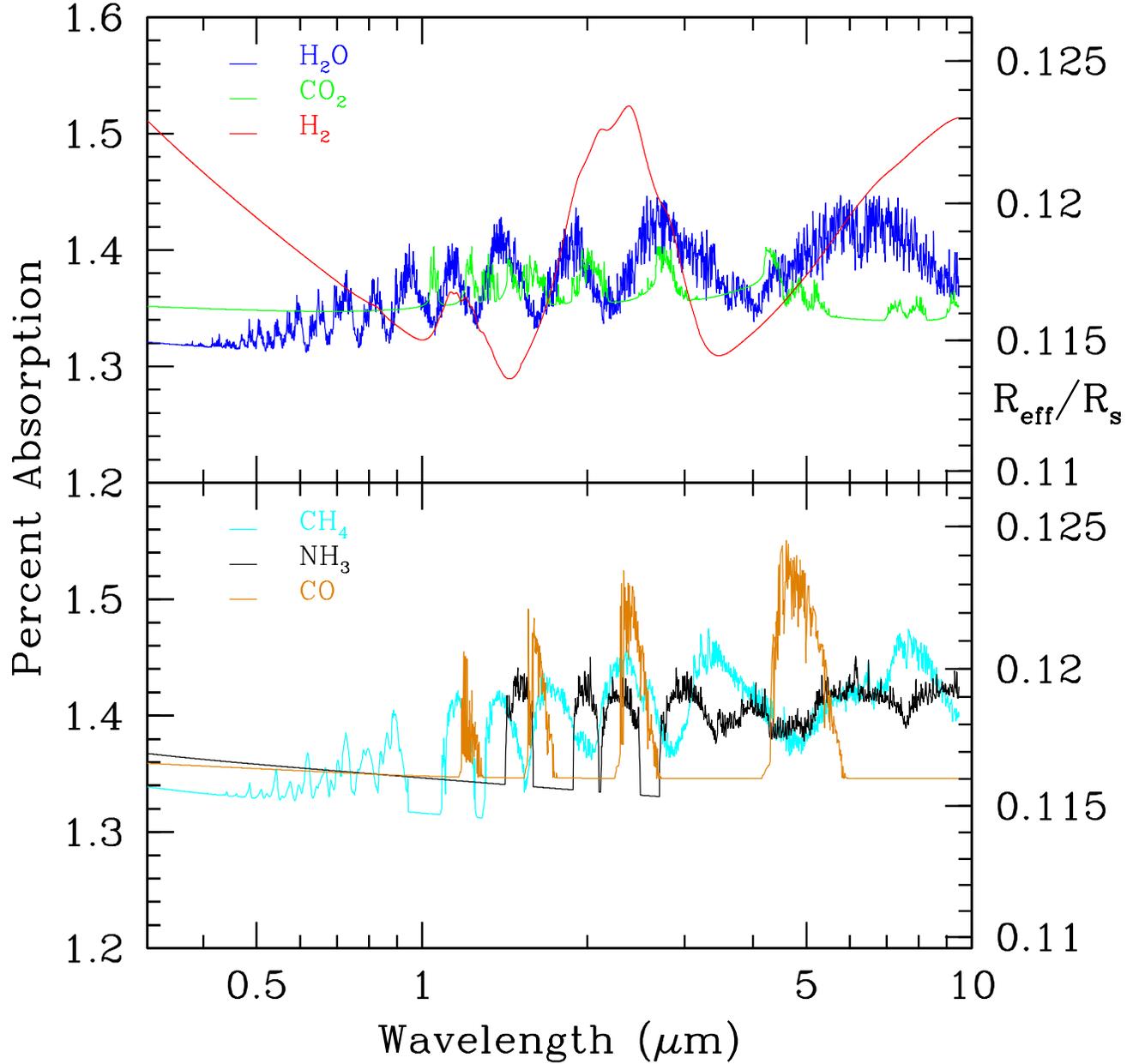} 
\end{center}
\caption{Transit depth in percent vs. wavelength in microns for generic transit spectra for hypothetical exoplanets with pure molecular atmospheres.  The scale heights of the atmospheres are inversely proportional to their molecular weights.  This proportionately affects the magnitudes of the spectral features, along with the slope of the Rayleigh continuum.  Smooth regions of the spectra represent the Rayleigh continuum, with the exception of ${\rm H_2 \, and \, CO_2}$, which additionally have smooth regions in their absorption spectra.  The spectra replicate the molecular absorption features shown in Figure \ref{opacities}.  Each spectrum was computed using the physical parameters and normalization for GJ 1214b.}
\label{pure}
\end{figure}

To test the code, we compute transit spectra for atmospheres with pure molecular compositions and for solar metallicities over a range of metallicities, temperatures, and opaque clouds placed at varying altitudes.  These calculations successfully reproduce the relations predicted by systematics, and they are discussed in detail in Section \ref{general}.  The spectra for pure molecular atmospheres succeed in reproducing the molecular opacity features of each species (compare Figures \ref{opacities} and \ref{pure}).  Smooth regions in the spectra represent the Rayleigh continuum (or, if the continuum is sufficiently transparent, the cloud tops or the planet's solid surface) at wavelengths where absorption is negligible.

We also compute spectra for GJ 1214b with compositions similar to those used in Miller-Ricci \& Fortney (2010) with a ``surface pressure'' of 1 bar.  At this pressure level, the atmosphere becomes opaque at all wavelengths in most models, so a deeper atmosphere will not affect the spectrum.  Our spectra successfully reproduce the molecular features of the Miller-Ricci \& Fortney (2010) computations.  They have systematically larger transit depths than Miller-Ricci \& Fortney (2010) in the molecular features by 10\%-20\%, and they agree to better than 10\% with the slope of the Rayleigh continuum.  The systematic offsets are most likely due to the use of different opacity tables.

\section{Generic Model Results}
\label{general}

For convenience, all generic spectra are computed using the physical parameters of GJ 1214b as given in Tables \ref{planets} and \ref{stars}, except as specified.  They are also normalized to the initial measurement of GJ 1214b: 1.35\% (13,500 ppm) in the MEarth band of 0.7-1.0 ${\rm \mu m}$ \citep{Charbonneau}.

\begin{figure}[htp]			
\begin{center}
\includegraphics[scale=0.9]{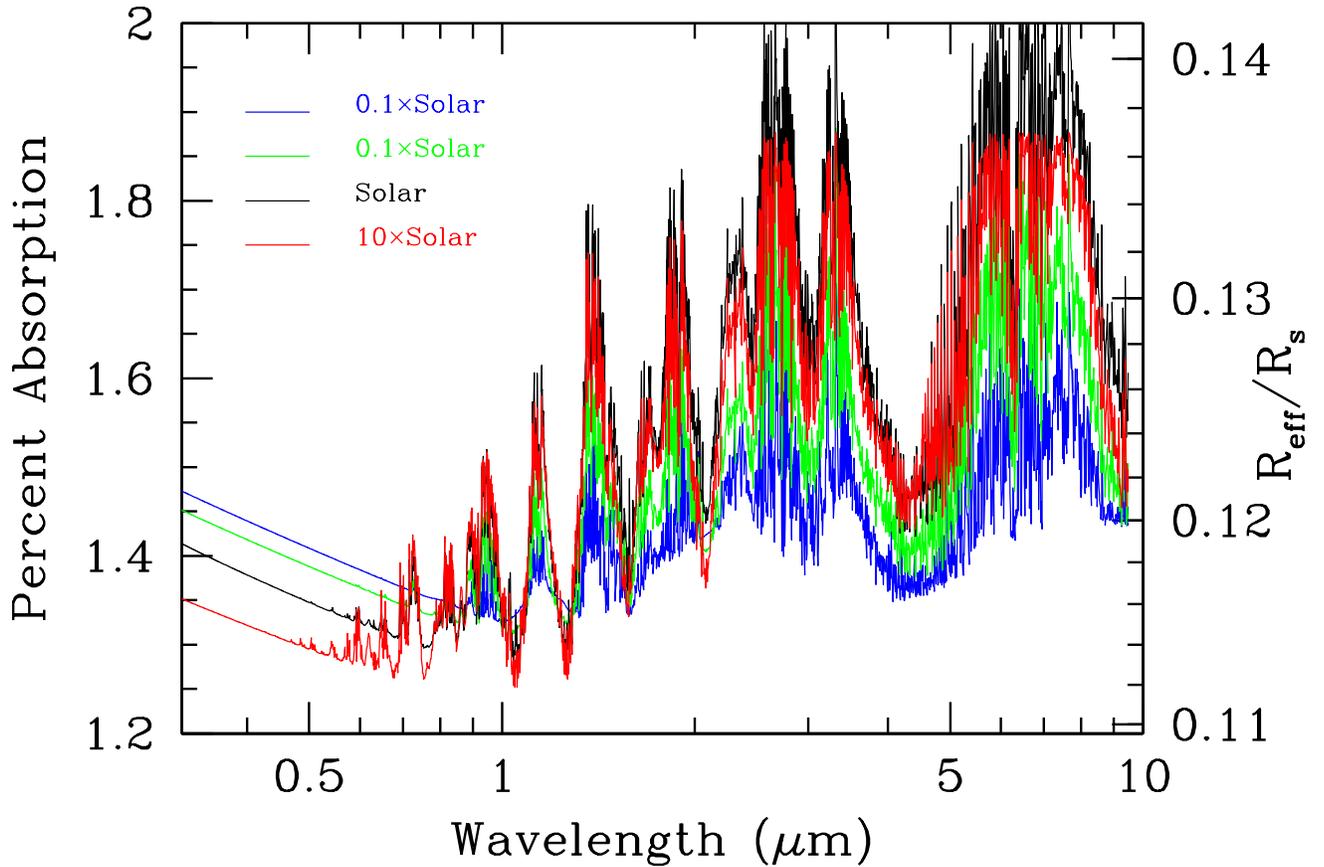} 
\end{center}
\caption{Transit depth in percent vs. wavelength in microns for generic transit spectra computed for exoplanets with hydrogen-rich atmospheres of varying metallicities.  For convenience, all of the generic spectra are computed using the physical parameters ($R = 2.678 R_\Earth, \, M = 6.55 M_\Earth, \, T = 470 {\rm K}$) of GJ 1214b, except as specified, and they are normalized to the initial measurement of GJ 1214b.  This normalization is 1.35\% (13500 ppm) absorption averaged over the MEarth band of 0.7-1.0 ${\rm \mu m}$ \citep{Charbonneau}.  Smooth sections of the graph represent the Rayleigh continuum.  The continuum falls with increasing metallicity because the normalization occurs in a band where scattering dominates at low metallicities, but absorption dominates at high metallicities, with very large absorption features forcing the continuum down.  The atmosphere is taken to be deep enough that the surface in obscured in transit at all wavelengths, and solar relative abundances and equilibrium chemistry are assumed.  Water and methane features are prominent.  Transit depth remains elevated between 2 and 3 ${\rm \mu m}$ because of a broad collision-induced absorption feature in hydrogen.}
\label{Metals}
\end{figure}

Generic spectra computed for hydrogen-rich atmospheres for which abundances are a fraction or multiple of solar are shown in Figure \ref{Metals}.  These spectra show transit depths increasing with metallicity.  This is expected since metals are the main source of extinction in a haze-free atmosphere.  However, for this normalization, the transit depth of the Rayleigh scattering tail decreases with increasing metallicity.  This is because of the normalization to the MEarth band, in which absorption dominates.  The spectral features in that band increase in amplitude with increasing metallicity relative to the Rayleigh continuum, so the continuum has a smaller relative transit depth at large metallicities.

\begin{figure}[htp]			
\begin{center}
\includegraphics[scale=0.9]{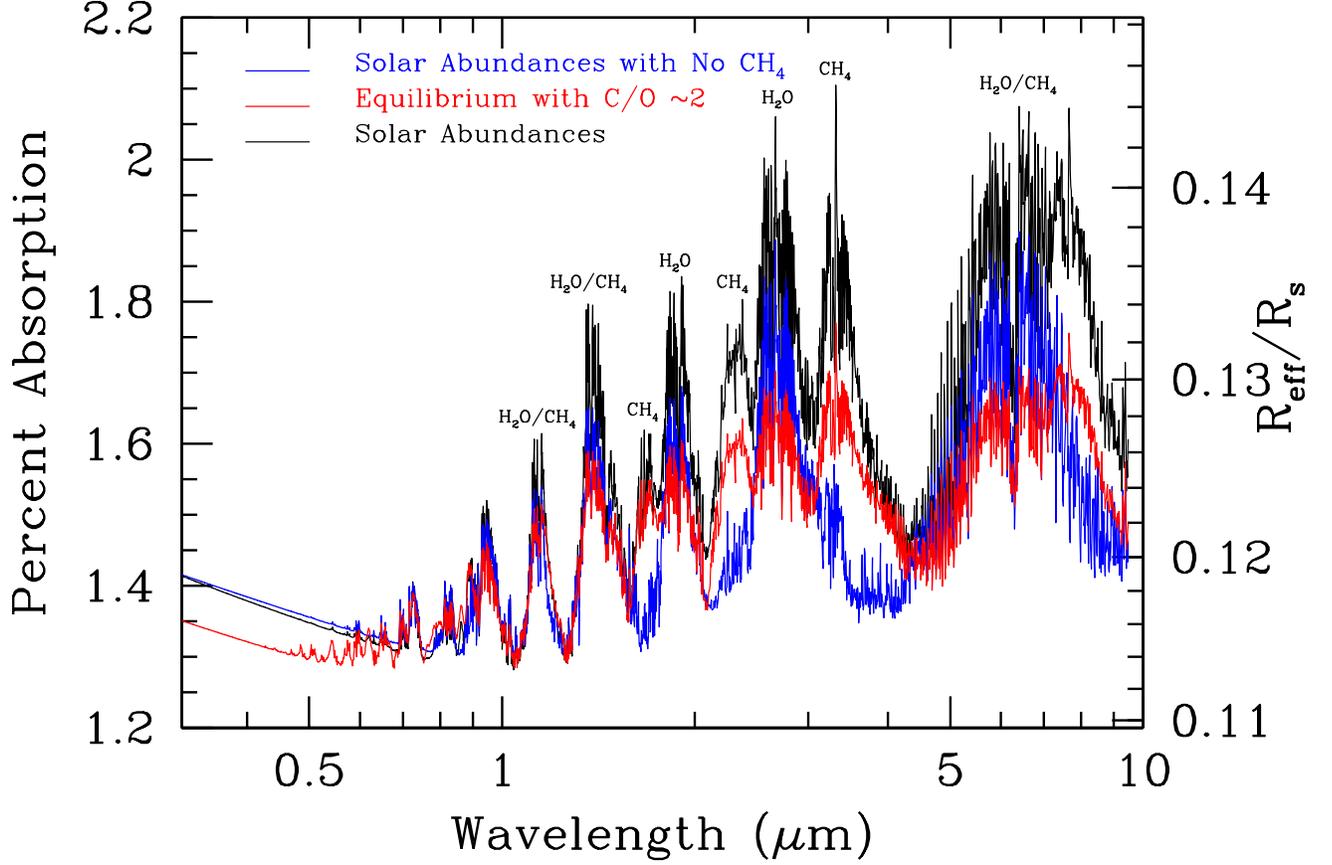} 
\end{center}
\caption{Transit depth in percent vs. wavelength in microns for generic transit spectra computed for exoplanets with hydrogen-rich atmospheres with non-solar relative abundances.  Blue: solar abundances with the abundance of methane set to zero.  The resulting spectrum is strongly water dominated.  Red: solar abundances with carbon and oxygen abundances exchanged, resulting in a C/O ratio of $\sim$2.  This causes water features to be muted and methane features to be enhanced.  The persistence of a large transit depth coinciding with the ``water feature'' at 2.7 ${\rm \mu m}$ is caused by a broad collision-induced absorption ${\rm H_2}$ feature.  The spectra are computed using the physical parameters and normalization for GJ 1214b.}
\label{Other}
\end{figure}

Two other hydrogen-rich atmosphere models are shown in Figure \ref{Other}.  Swapping the abundances of carbon and oxygen from solar levels--in effect, raising the C/O ratio from 0.5 to 2.0--diminishes the spectral features of water and enhances those of methane.  Some of these features overlap and, thus, those wavelengths are not strongly affected.  The greater extinction opacity of methane at near-infrared wavelengths causes the Rayleigh continuum to be normalized to a lower transit depth.  For the model with no methane, the spectrum is more strongly dominated by water features, and the normalization is not strongly affected.

\begin{figure}[htp]			
\begin{center}
\includegraphics[scale=0.9]{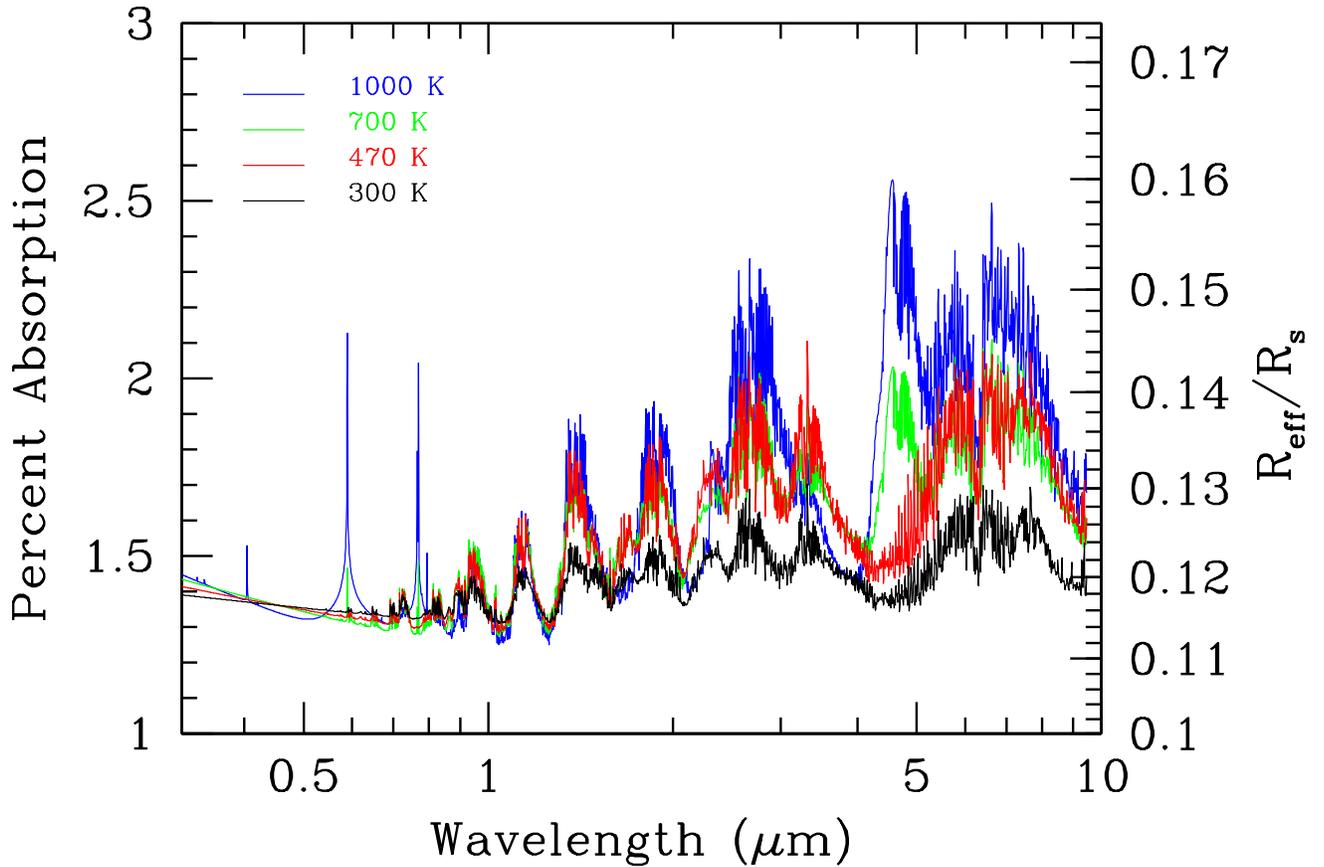} 
\end{center}
\caption{Transit depth in percent vs. wavelength in microns for generic transit spectra computed for exoplanets with solar-abundance atmospheres at various equilibrium temperatures.  Higher temperatures result in larger scale heights and larger transit depths.  The slope of the Rayleigh continuum also increases with temperature due to this effect.  At and above 700 K, the methane abundance drops, resulting in weaker overlapping methane and water features (seen blueward of 1.7 ${\rm \mu m}$) and much weaker isolated methane features (seen redward of 1.7 ${\rm \mu m}$), leaving a more water-dominated spectrum.  The 4.7 ${\rm \mu m}$ CO line and narrow alkali metal lines begin to appear at 700 K and become very prominent by 1000 K.  The spectra are computed using the physical parameters and normalization for GJ 1214b.}
\label{Temperature}
\end{figure}

The effects of temperature on the generic transit spectra are shown in Figure \ref{Temperature}.  A higher temperature results in a larger scale height, and, as expected, a larger continuum slope at short wavelengths.  As the temperature increases, two major chemical changes are evident as well--the suppression of methane and the appearance of strong alkali metal lines.

\begin{figure}[htp]			
\begin{center}
\includegraphics[scale=0.9]{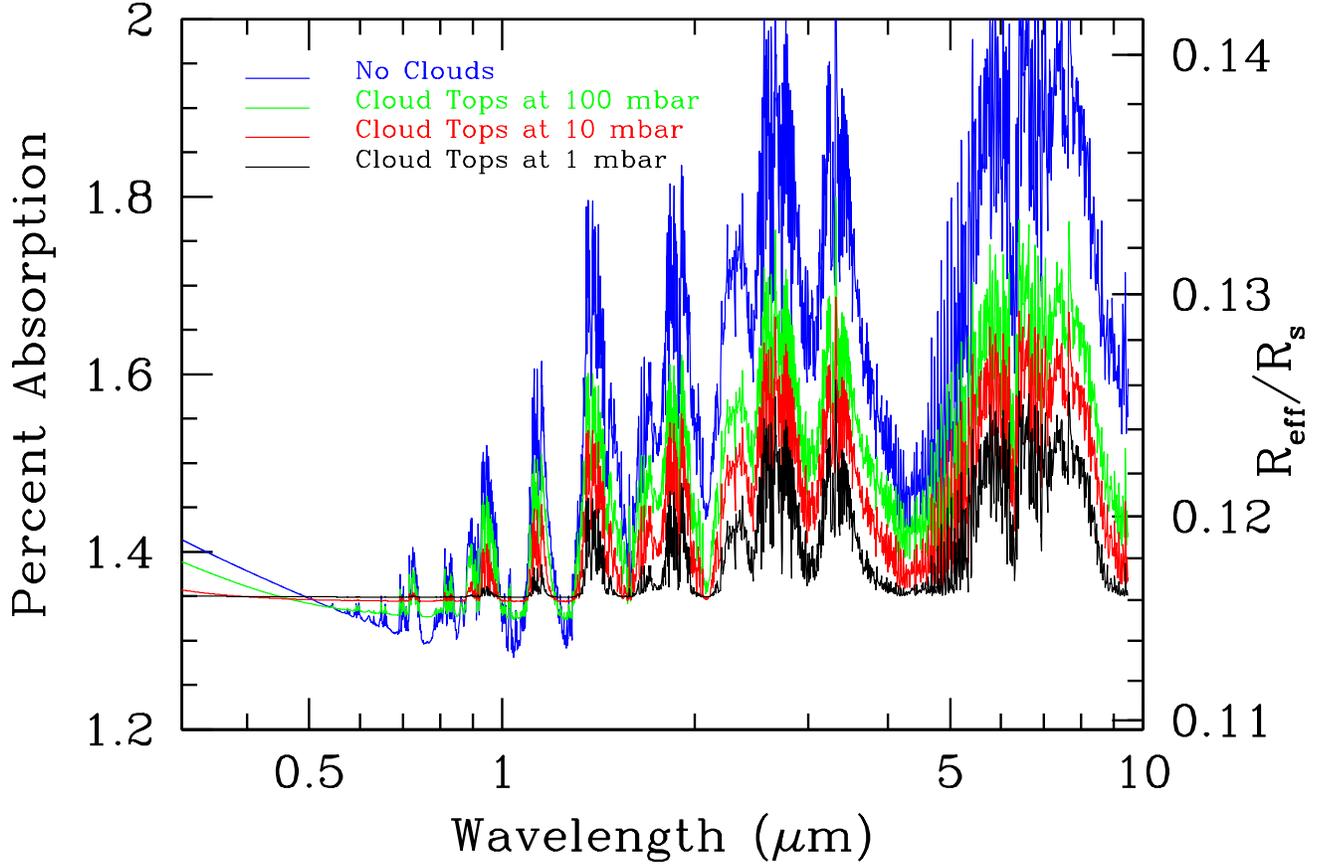} 
\end{center}
\caption{Transit depth in percent vs. wavelength in microns for generic transit spectra computed for exoplanets with solar-abundance atmospheres with achromatically opaque cloud decks at various pressure levels.  The 1 bar level represents a cloudless atmosphere, that is, any clouds are deep enough that they are obscured by molecular absorption and scattering.  Horizontal smooth regions indicate wavelengths where the atmosphere is transparent down to the cloud deck, so that the planet presents a fixed radius in transit with respect to wavelength.  With very high cloud decks, many features are smoothed out, and those that remain are much weaker.  If the clouds are above the 10-mbar level, they also obscure the Rayleigh continuum, leaving it horizontal.  The slope of the Rayleigh continuum decreases gradually before this point, since the clouds partially and achromatically obscure it.  The spectra are computed using the physical parameters and normalization for GJ 1214b.}
\label{Pressure}
\end{figure}

An opaque cloud layer can suppress the molecular spectral features if it lies at high enough altitudes.  If light is not sufficiently blocked as it grazes the cloud tops, the observed radius should be very near the cloud tops in that wavelength.  This effect is demonstrated in Figure \ref{Pressure}, which includes spectra with cloud layers at different pressure levels.  Flat, horizontal sections of the spectra represent wavelengths where the atmosphere is transparent down to the cloud tops.  If the clouds are deeper than the 1 bar level, they are usually obscured at all wavelengths, so higher pressures have no effect on the spectrum.  However, for some atmospheres composed of pure molecular species, such as carbon dioxide, light can penetrate deeper at some infrared wavelengths, where both absorption and scattering are negligible, so these flat regions due to clouds can persist to very high pressures in such an atmosphere.


Models for pure atmospheric compositions are shown in Figure \ref{pure}, to be compared with Figure \ref{opacities}, which depicts the absorption opacities for the same species at a fixed temperature and density.  Nitrogen is assumed to have Rayleigh scattering opacity only, as are the other species at wavelengths outside their respective absorption bands.  All of these results are consistent with the expected dependencies on physical and atmospheric parameters and, thus, provide a further validation of our methodology.

\begin{figure}[htp]			
\begin{center}
\includegraphics[scale=0.9]{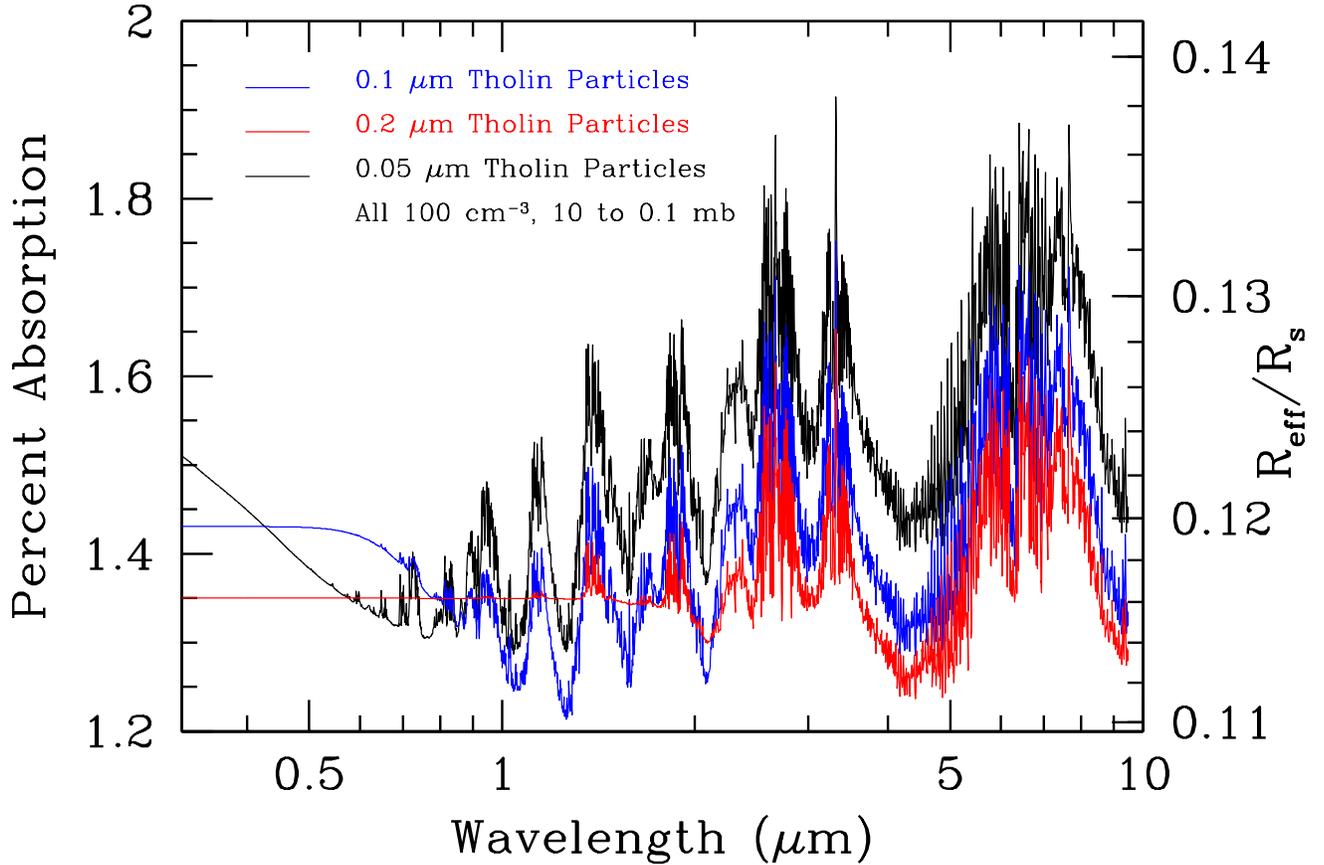} 
\end{center}
\caption{Transit depth in percent vs. wavelength in microns for generic transit spectra computed for exoplanets with solar-abundance atmospheres with tholin haze layers with different particle sizes.  In each case, the haze has a number density of 100 cm$^{-3}$ and extends from the 10 mbar pressure level to the 0.1 mbar pressure level.  Horizontal smooth regions indicate wavelengths where the haze layer is opaque and the atmosphere is nearly transparent above it, so that the planet presents a fixed radius in transit with respect to wavelength.  Sloped smooth regions correspond to wavelengths in which the haze layer is not opaque at the cloud tops, but it still obscures the absorption features.  The slopes are due to the wavelength dependence of Mie scattering.  The spectra are computed using the physical parameters and normalization for GJ 1214b.}
\label{Haze}
\end{figure}

We add haze to the atmosphere using the Mie approximation for spherical particles.  We find that dense hazes ($n \gtrsim 10^{5} \, \mathrm{cm^{-3}}$ for most typical sizes) are essentially opaque in the wavelength range we study and result in flat regions in the spectrum.  However, very thin hazes ($n \lesssim 10^2 \, \mathrm{cm^{-3}}$ for typical particle sizes) become transparent at some wavelengths.  In general, these thin hazes tend to produce sloped regions of the spectrum similar to those produced by Rayleigh scattering.  However, these regions do not span a wide range of transit depths because many molecular features persist to much lower pressures than typical hazes.  Examples of spectra with haze included are shown in Figure \ref{Haze}.

\section{GJ 1214b}
\label{gj1214}

\subsection{Prior Observations}
\label{prior}

\begin{figure}[htp]			
\begin{center}
\includegraphics[scale=0.9]{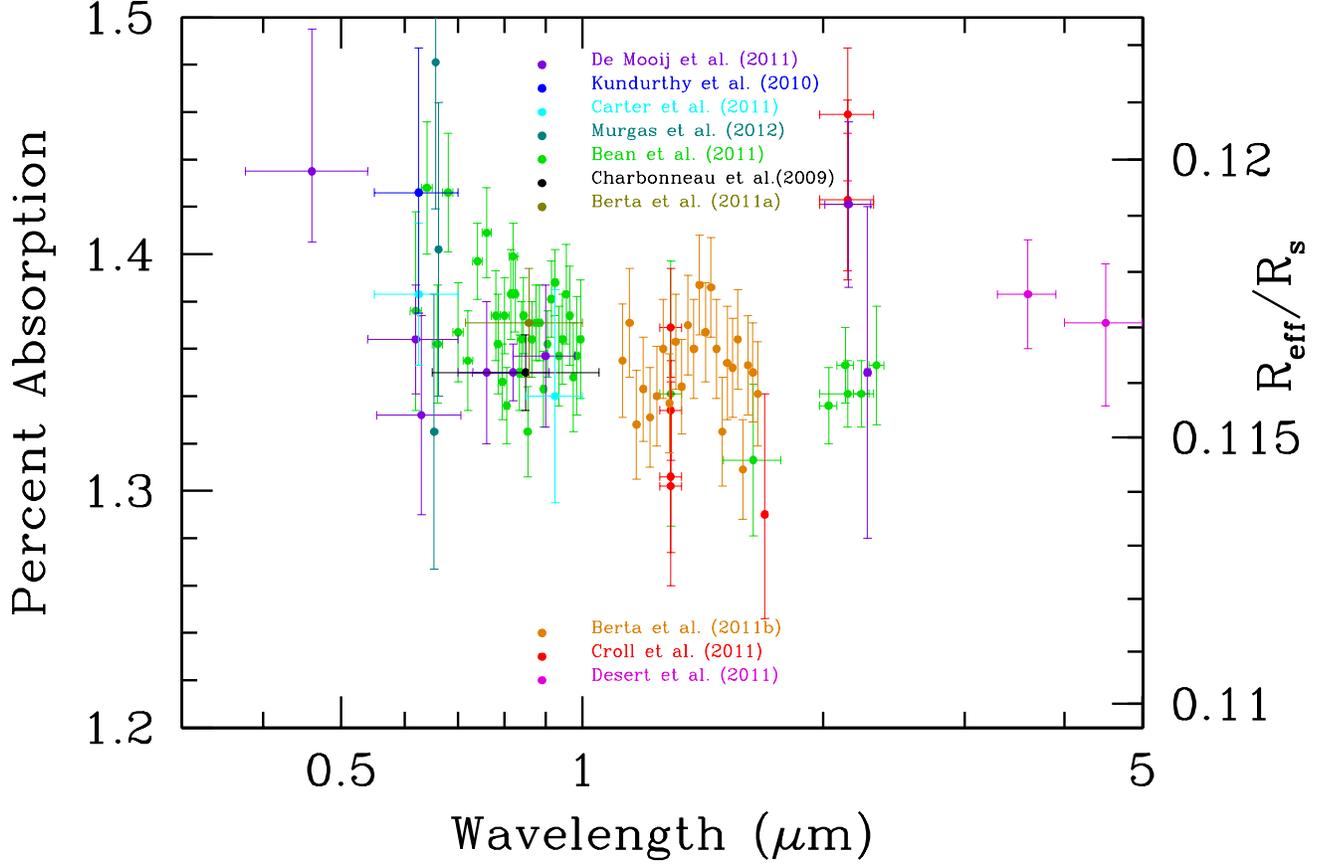} 
\end{center}
\caption{Published photometric and spectroscopic transit data for GJ 1214b: transit depth in percent vs. wavelength in microns.  Data from Sada et al. (2010) have been omitted due to large uncertainties.  Data from Crossfield et al. (2011) have been omitted because they present a relative spectrum rather than an absolute percentage of absorption.  All other published data are shown with 1-$\sigma$ uncertainties.  The horizontal ``error bars'' are not uncertainties, but instead represent the widths of the observational bands.}
\label{Data}
\end{figure}

The photometric and spectroscopic data for GJ 1214b available at this writing are shown in Figure \ref{Data}.  D\'{e}sert et al. (2011), working in the Warm Spitzer 3.6 and 4.5 ${\rm \mu m}$ bands, observed a flat transit spectrum of GJ 1214b and concluded that its atmosphere must have a small scale height and, thus, a high molecular weight, probably with a large proportion of water.  Bean et al. (2010), working in the $I$ band, came to the same conclusion, although they also suggested that the planet could have a layer of clouds or hazes above the 300-mbar level.

This picture changed when Croll et al. (2011) observed a deep $K_s$-band transit, but shallow $J$ and $\mathrm{CH_4On \, (1.69 \, \mu m)}$ band transits,  suggesting a hydrogen-rich, low-molecular-weight atmosphere, observable in the $K_s$ band, that included a haze layer to mute features at other wavelengths.  The anomalous $K_s$-band feature was attributed to either water or methane.

However, Crossfield et al. (2011), based on near-IR spectroscopy, argue that a hydrogen-rich atmosphere is not favored, and that if a large-scale-height atmosphere were present, it must be out of chemical equilibrium (e.g., depleted in methane) or muted by a haze or cloud layer.  Bean et al. (2011) revised their earlier assessment with data in the $R$, $J$, $H$, and $K$ bands.  These data support a high-molecular-weight atmosphere, particularly with shallow K-band transits.  However, their reported results in the R-band show some evidence for a Rayleigh scattering tail at short wavelengths, with a slope consistent with a hydrogen-rich atmosphere.

The most recent observations continue to give conflicting results.  De Mooij et al. (2011), taking measurements in a wide range of visible and infrared bands, obtained results that reinforce both the conclusion of both apparent Rayleigh scattering at short wavelengths in Bean et al. (2011) and Croll's deep $K_s$-band measurement, providing further evidence for a hydrogen-rich atmosphere.  However, the small relative size of the observed $K_s$-band feature, compared with theoretical predictions for a hydrogen-rich atmosphere, suggested an improbably low metallicity for the planet, despite orbiting a star with normal metallicity \citep{de Mooij}.  Finally, Berta et al. (2011b) observed what appears to be a feature at $\sim$1.5 ${\rm \mu m}$ with the {\it Hubble Space Telescope's} WFC3, which could correspond to water, the depth of which is consistent with a high-molecular-weight atmosphere.

To explain these conflicting results, one of the most widely postulated models is a hydrogen-rich atmosphere with a layer of clouds or haze near or above the 100 mbar pressure level \citep{Croll,Bean2011}, or above the 10 mbar level in Berta et al. (2011b).  Haze has also been suggested to explain the weak alkali metal features in transits of the giant planet HD 189733b \citep{Pont}.  The other alternative is a high-molecular-weight atmosphere with a high water content \citep{Desert,Berta II}.

\subsection{Models without Haze}
\label{nohaze}

Our full list of atmosphere models for GJ 1214b is given in Table 3.  Assuming that GJ 1214b has a low albedo and its heat flow is distributed uniformly over its surface, its equilibrium temperature is $\sim$555 K \citep{Miller-Ricci2010}.  If we further assume that the characterization of a Rayleigh scattering tail is correct ($\alpha = 4$), then the complete data set suggests a mean molecular weight for GJ 1214b's atmosphere of around 2.7, based on the analytic model given in the Appendix.  This mean molecular weight definitely represents a hydrogen-rich atmosphere, but it leaves the problem of the lack of expected spectral features unresolved.

\begin{figure}[htp]			
\begin{center}
\includegraphics[scale=0.9]{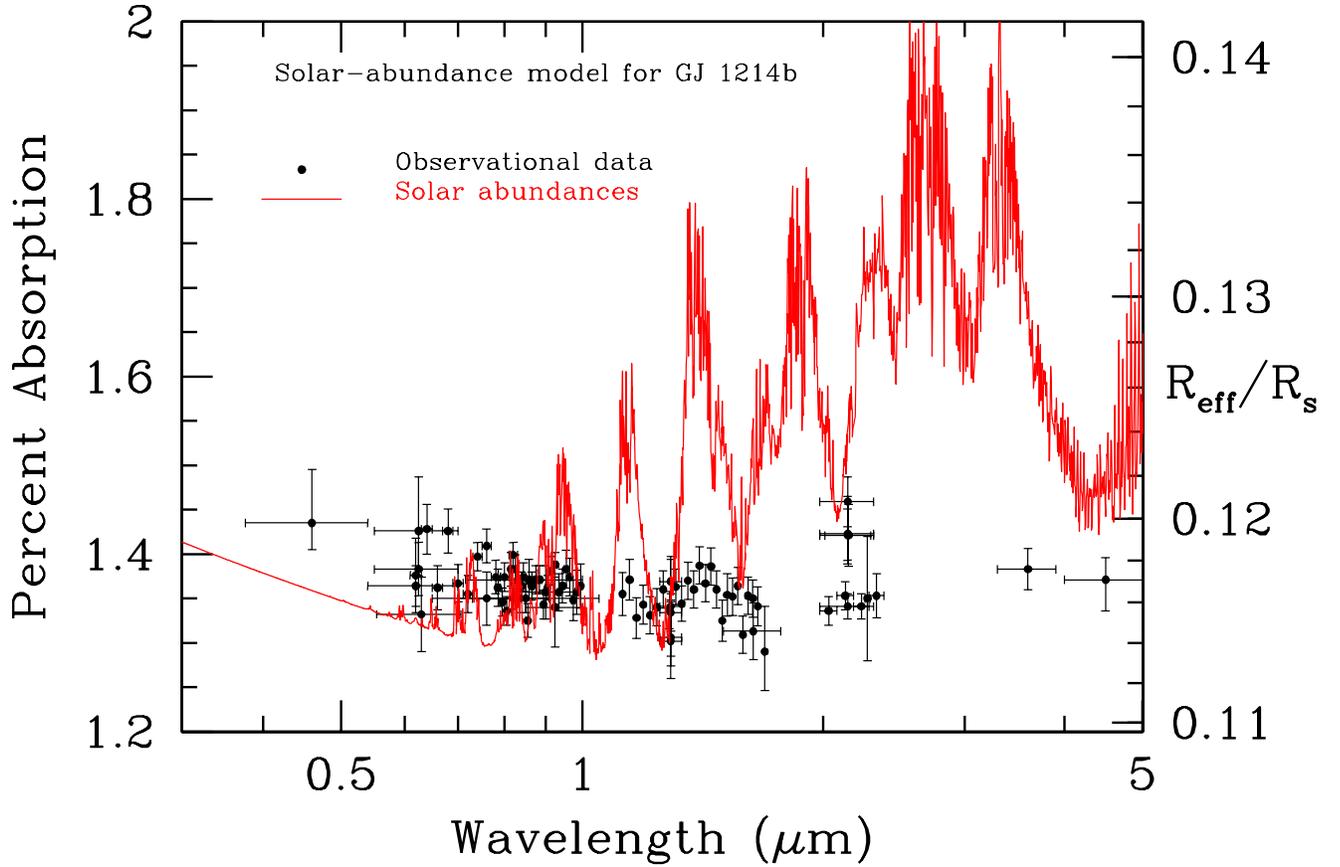} 
\end{center}
\caption{Transit depth in percent vs. wavelength in microns for a solar-abundance atmospheric model for GJ 1214b compared with the observational data.  The spectrum is normalized to the initial observation of GJ 1214b of a transit depth of 1.35\% (13,500 ppm) in the MEarth band pass of 0.7-1.0 ${\rm \mu m}$ \citep{Charbonneau}.  The spectral features are far too large to fit the data.  Water features are most prominent, and one of them, at 1.5 ${\rm \mu m}$, corresponds in wavelength to one of the observed features.  Distinct methane features occur at 2.2 and 3.3 ${\rm \mu m}$, with most of the others overlapping with water features.  Collision-induced absorption by hydrogen also contributes to the large transit depth between 2 and 3 ${\rm \mu m}$.  The smooth portion of the spectrum represents the Rayleigh continuum.  It has a similar slope to the rise at short wavelengths in the observations, but occurs only at shorter wavelengths, since it is dominated by absorption elsewhere.}
\label{solar-model}
\end{figure}

Working from this initial prediction of a hydrogen-rich atmosphere, Figure \ref{solar-model} shows the spectrum of a solar-abundance atmosphere model for GJ 1214b compared with the observational data.  It is normalized to the initial observation of GJ 1214b of a transit depth of 1.35\% (13500 ppm) in the ``MEarth'' bandpass of 0.7-1.0 ${\rm \mu m}$ \citep{Charbonneau}.  This model spectrum bears almost no resemblance to the data, since the features are far too large and of different relative proportions than the observed features.  Water features are most prominent in the solar model spectrum, with one that coincides with the suspected 1.5 ${\rm \mu m}$ feature, except in magnitude.  The Rayleigh continuum is visible as a smooth rise in transit depth blueward of 0.7 ${\rm \mu m}$ with a similar slope to the observed rise at short wavelengths, but a far lower transit depth.  Also, absorption dominates over scattering well into the optical range, so that the Rayleigh tail is predicted primarily at wavelengths shorter than the observations.  Methane features can also be identified, such as those at 2.2 and 3.3 ${\rm \mu m}$, although the prominent methane features at shorter wavelengths overlap with water features.

By ignoring one or more observed features at a time, the small amplitude of the remaining features can be explained by one or more of four possibilities: (1) the atmosphere contains low abundances of molecules with large opacities; (2) the atmosphere contains an opaque cloud layer that mutes the molecular absorption features; (3) the atmosphere contains a translucent haze layer that mutes the molecular absorption features; and (4) the atmosphere has a high molecular weight, which results in a smaller scale height, a shallower atmosphere, and, thus, a smaller variation in transit depth.

\begin{figure}[htp]			
\begin{center}
\includegraphics[scale=0.9]{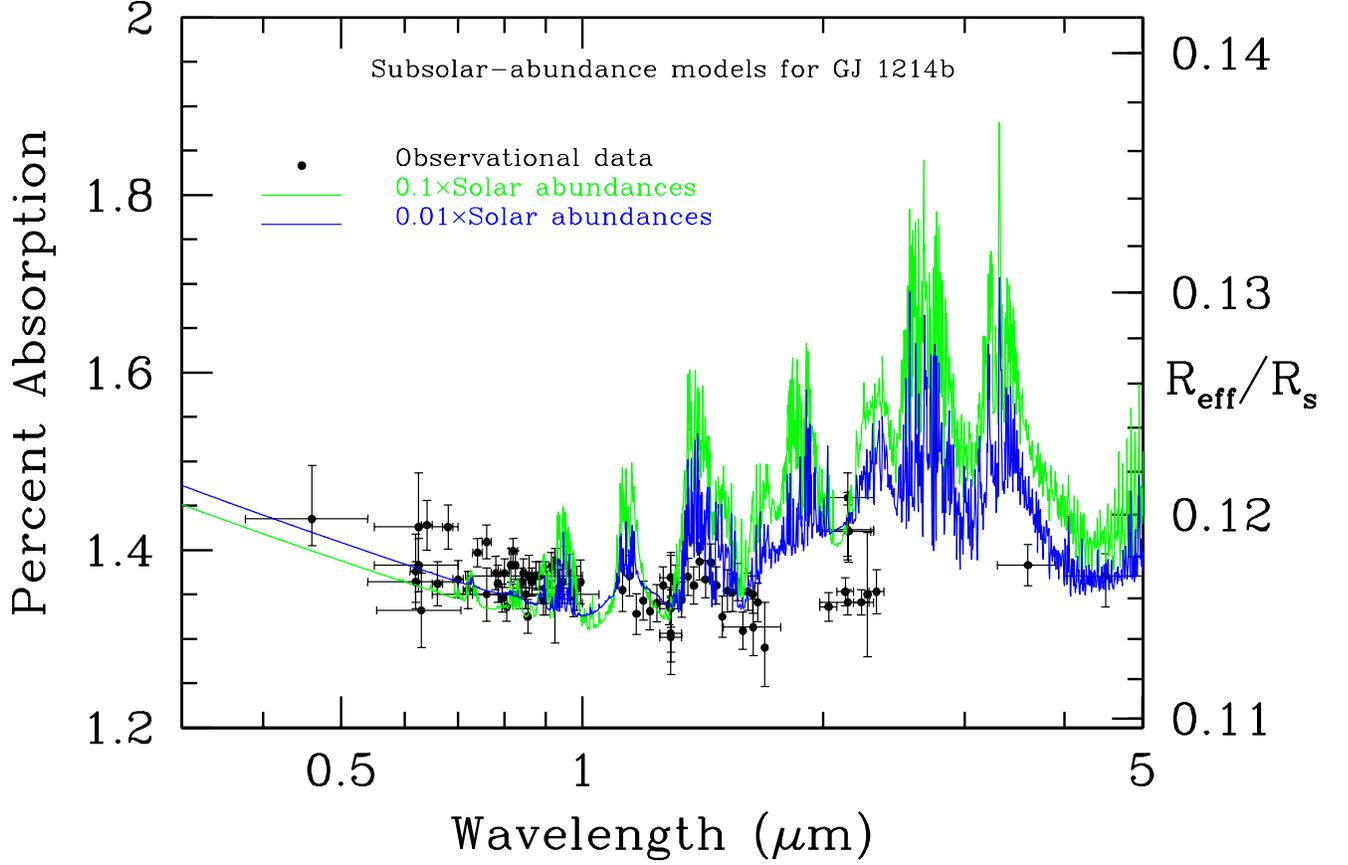} 
\end{center}
\caption{Transit depth in percent vs. wavelength in microns for 0.1$\times$ solar and 0.01$\times$ solar abundance atmosphere models for GJ 1214b.  Hydrogen has a relatively low opacity compared with metal compounds, so reducing the metal abundances results in smaller spectral features.  At ${\rm -2<[Fe/H]<-3}$, the spectrum is consistent with most of the infrared observations, including the suggested 1.5 ${\rm \mu m}$ feature, the large transit depth observed in the $K_s$-band, and the mid-infrared Spitzer observations.  Furthermore, the computed Rayleigh tail extends redward into the observed tail and has a similar transit depth to observations, unlike in the solar-abundance model.  This is a relatively good fit, but we do not support it because it conflict's with GJ 1214's observed metallicity of ${\rm [Fe/H]} = +0.39$ \citep{Charbonneau}.}
\label{metals-model}
\end{figure}

Figure \ref{metals-model} shows the spectrum of two low-metallicity hydrogen-rich models.  Hydrogen has a relatively low absorption opacity compared with  metal compounds, so reducing the metal abundance makes most of the features much smaller in amplitude.  The model spectrum begins to conform with the data at metallicity of ${\rm [Fe/H]} \lesssim -2$; the computed Rayleigh tail extends redward into the observed tail, and the spectrum is at least marginally consistent with most of the infrared observations, including the suggested 1.5${\rm \mu m}$ feature, the $K_s$-band, and the Spitzer bands.  While this is a relatively good fit, given the difficulty of reconciling the various observations at different wavelengths, the star GJ 1214 has a metallicity of ${\rm [Fe/H]} = +0.39$ \citep{Charbonneau}, making the existence of an associated planet with such a low metallicity extremely unlikely.  We therefore do not advance this solution.

\begin{figure}[htp]			
\begin{center}
\includegraphics[scale=0.9]{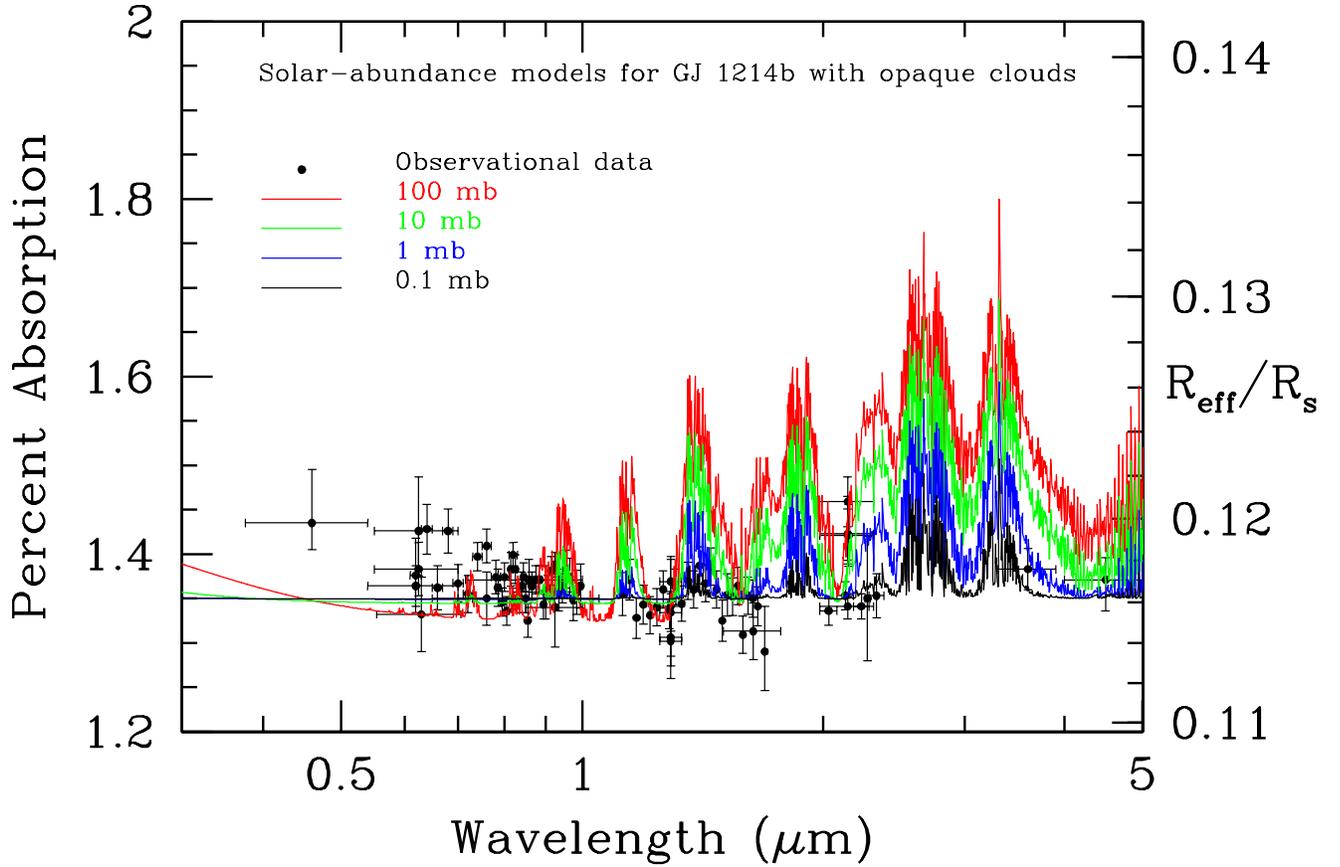} 
\end{center}
\caption{Transit depth in percent vs. wavelength in microns for atmosphere models for GJ 1214b with solar abundances and an opaque cloud layer at various pressure levels.  If the cloud tops are near or above the 1 mbar level, the spectrum is flat enough to become consistent with observations.  In this case, only very large features should be seen in transit in the rarefied regions above the cloud layer.  Specifically, the spectrum is consistent with most of the infrared features (the $K_s$ band being the primary exception).  It would be a viable model for GJ 1214b if only low-transit-depth datapoints are considered.  Unfortunately, placing clouds at such a high altitude also suppresses the Rayleigh tail in addition to the molecular features, leaving the spectrum inconsistent with the observed short-wavelength rise.}
\label{clouds-model}
\end{figure}

Figure \ref{clouds-model} portrays spectra of model atmospheres with solar abundances with the addition at an opaque cloud layer at various pressure levels.  An opaque cloud layer near or above the 100-mbar level has been suggested in several other models for GJ 1214b \citep{Croll,Bean2011}, and above the 10 mbar level in Berta et al. (2011b).  We find that the spectrum becomes flat enough to be consistent with observations only if the cloud layer cuts off near or above the 1 mbar level.  In this case, only very large features remain visible in transit in the rarefied regions above the cloud layer.  If the cloud tops are between 1 mbar and 0.1 mbar, the spectrum is consistent with most of the measured infrared features (the ${\rm K_s}$ band being the primary exception) and would be a viable model if all of the high-amplitude features are rejected.  Unfortunately, placing clouds at such a high altitude also suppresses the Rayleigh tail due to the molecular features, leaving the spectrum inconsistent with the observed short-wavelength rise.  However, a translucent haze layer remains a viable alternative (see below).

Figure \ref{water-model} provides spectra of model atmospheres with high molecular weights, consisting of various proportions of water, with the remainder nitrogen, which we take to have only Rayleigh scattering opacity.  With a higher molecular weight, the scale height of the atmosphere is smaller, as are the spectra features.  Almost any proportion of water is consistent with most of the observed infrared features, again with the $K_s$ band being the primary exception.  The best fit is an abundance of $\lesssim 10\%$, water, which reproduces the suggested 1.5 ${\rm \mu m}$ feature in Berta et al. (2011b) and the K-band measurements of Bean et al. (2011).  However, these models fail to produce the short-wavelength rise, and the lower abundances fail to produce the slight observed rise in the mid-infrared \citep{Desert}.  Furthermore, the molecular weights of water and nitrogen result in a scale height so small that the slope of the Rayleigh tail is much less than that of the observed rise.  This can be confirmed using the analytic model in the Appendix.  The slope of the Rayleigh tail in a pure water atmosphere (molecular weight 18) is expected to be less than one-sixth the slope of the observed rise.

\begin{figure}[htp]			
\begin{center}
\includegraphics[scale=0.9]{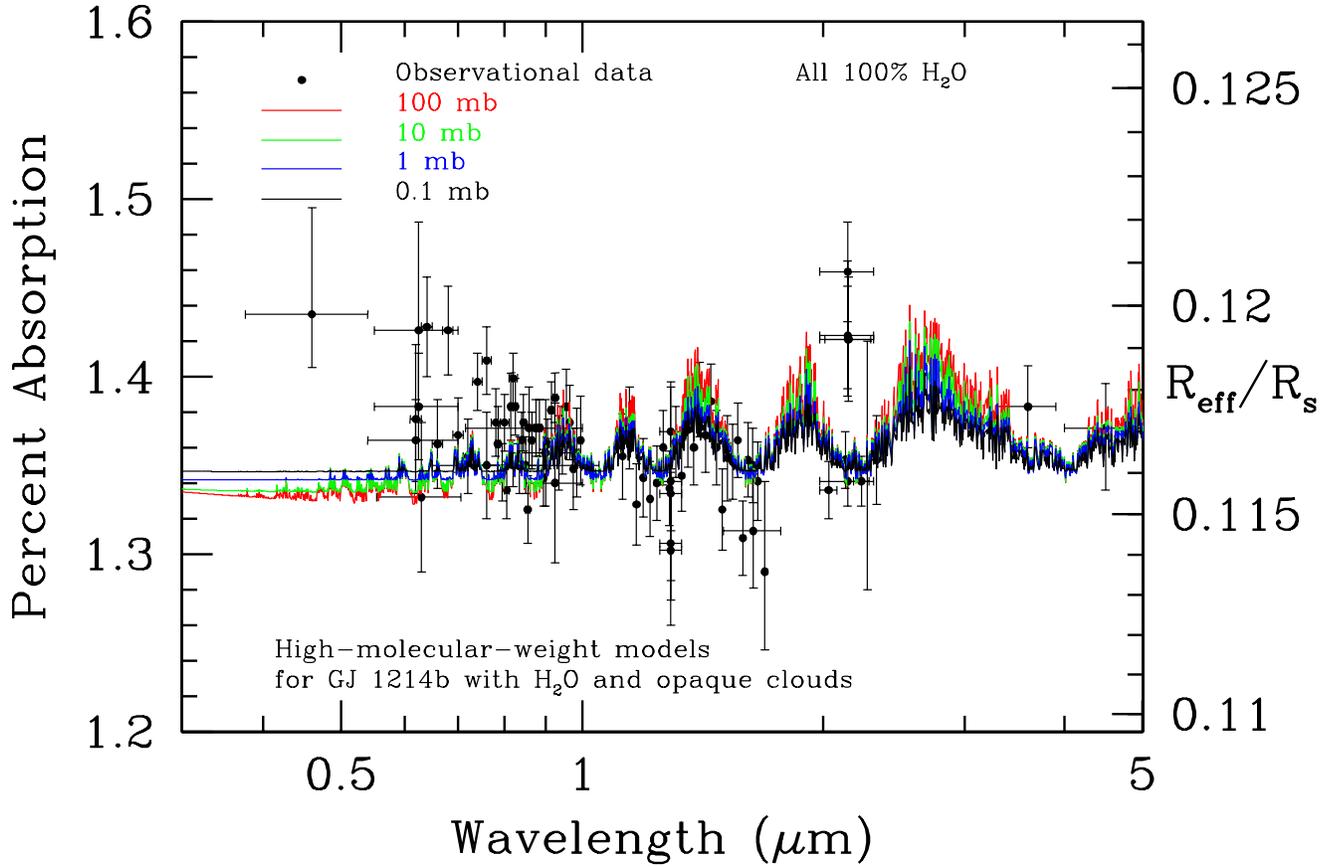} 
\end{center}
\caption{Transit depth in percent vs. wavelength in microns for pure-water atmosphere models for GJ 1214b with opaque clouds at various pressure levels.  The higher abundance of water compared to the hydrogen-rich models results in larger absorption at low pressures above the cloud tops and, thus, less suppression of absorption features.  All of these models are consistent with the suspected 1.5 ${\rm \mu m}$ feature and the $K$-band observations.  However, the clouds suppress the Rayleigh tail, leaving it flat or nearly flat.  Also, the spectra do not agree with the mid-infrared features.}
\label{water-clouds-model}
\end{figure}

As before, the addition of an opaque cloud layer only exacerbates the problem.  This effect is illustrated in Figure \ref{water-clouds-model}, specifically, with a pure water atmosphere.  As before, the addition of clouds improves the agreement with the observed molecular features, but obscures the Rayleigh tail, leaving it a flat continuum if the cloud tops are above 10 mbar.

\begin{figure}[htp]			
\begin{center}
\includegraphics[scale=0.9]{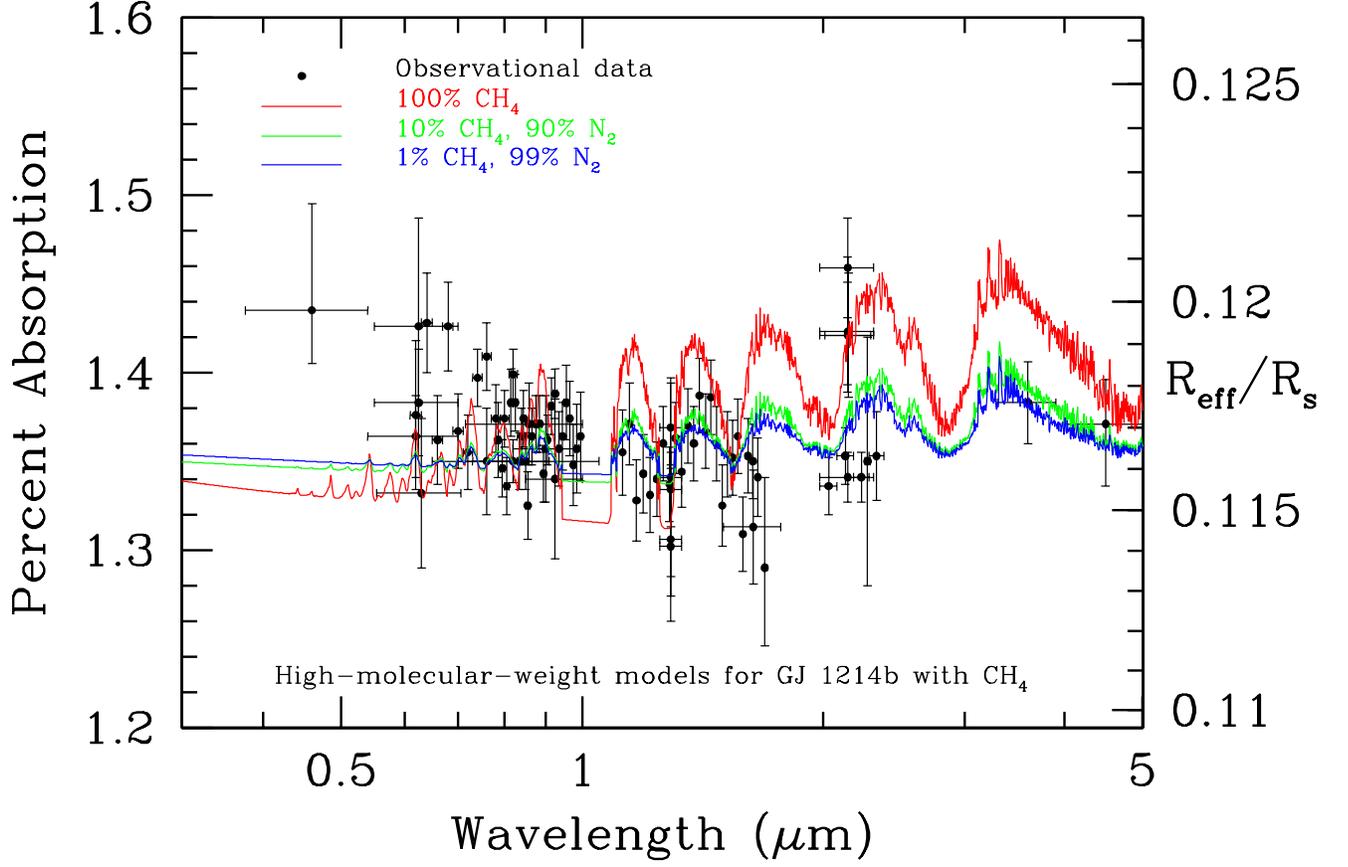} 
\end{center}
\caption{Transit depth in percent vs. wavelength in microns for atmosphere models of GJ 1214b with a solar-abundance atmosphere with various proportions of methane with the remainder nitrogen, which we take to have only Rayleigh scattering opacity.  Pure methane reproduces the large $K_s$-band observations, but also predicts features that are too large in amplitude in other near- and mid-infrared wavelengths.  Alternatively, an abundance of $\lesssim$10\% methane fits the suspected 1.5 ${\rm \mu m}$ feature well, but predicts additional near-infrared features that are not observed.  Thus, we do not advance methane as an alternative to water to explain the observed absorption features.}
\label{ch4}
\end{figure}

Of the other molecules we consider, methane is the only one that might be consistent with the observations (Figure \ref{ch4}).  A pure methane atmosphere fits the large transit depth observed in the $K_s$ band, but the overall fit is poor.  A composition of $\lesssim 10\%$ methane does better, fitting the mid-infrared datapoints and the 1.5-${\rm \mu m}$ feature well, but it predicts additional near-infrared features that are not observed, and it does not fit either of the conflicting $K$-band and $K_s$-band measurements.  Therefore, we conclude that the principal source of absorption opacity is most likely water.

\subsection{Models with Haze}
\label{haze}

\begin{figure}[htp]			
\begin{center}
\includegraphics[scale=0.9]{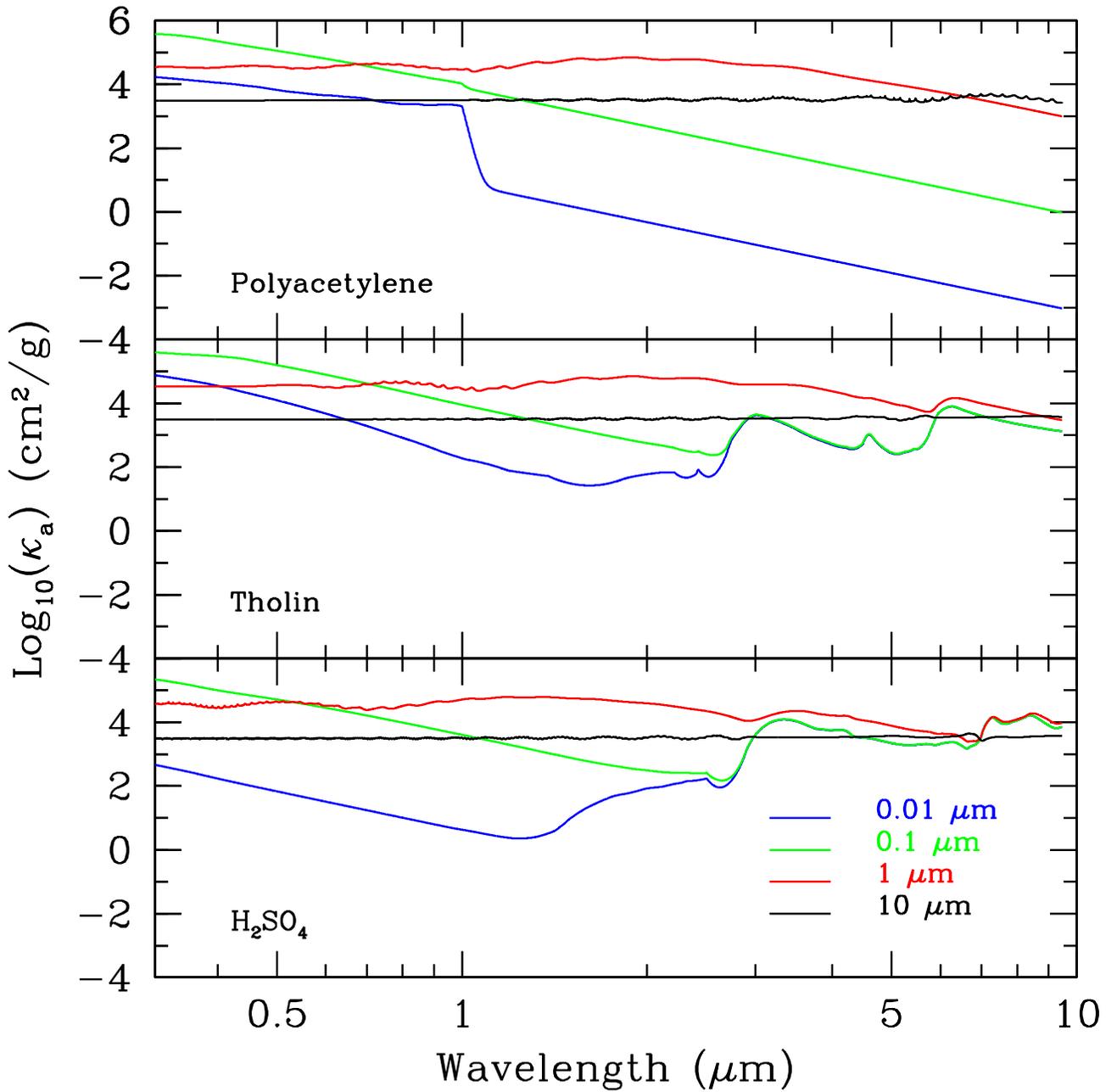} 
\end{center}
\caption{Extinction opactities in ${\rm cm^2 \, g^{-1}}$ of polyacetylene, tholin, and sulfuric acid hazes of various monodispersed particle sizes vs. wavelength in microns.  We have assumed all of the haze species have a density of 1 ${\rm g \, cm^{-3}}$ and that they are the only contributor to atmospheric extinction.  Thus, they are proportional to the extinction cross section of the particles multiplied by the number density and divided by the mass density.  Constant-sloped sections indicate Rayleigh-like ``$\lambda^{-4}$'' behavior occurring at particle sizes significantly smaller than the wavelength.  For polyacetylenes, this behavior occurs at long wavelengths.  For sulfuric acid, it occurs at shorter wavelengths, with other wavelength-dependent behavior dominating in the mid-infrared.  For tholins, similar spectral features dominate over Rayleigh-like behavior at most wavelengths.  The weak oscillatory features are artifacts of the use of monodispersed particle distributions.}
\label{haze-xsec}
\end{figure}

On the problem of the short-wavelength rise, the high-molecular-weight model could possibly be salvaged with the addition of a haze layer.  Some possible haze components in this temperature range are polyacetylene, tholin, and sulfuric acid.  Sulfuric acid is known to be in the atmosphere of Venus, but is considered less probable than carbon-based haze species, most notably due to the uncertainty of producing it in a water- or hydrogen-rich atmosphere.  For this reason, and because its extinction opacity has a similar wavelength dependence to tholins at most wavelengths and particle sizes (see Figure \ref{haze-xsec}), we do not consider it as a solution for GJ 1214b in this work.  However, we do produce some general models including sulfuric acid haze in our online library.  These similarities in behavior also mean that the haze composition is ambiguous without more data.

Figure \ref{haze-xsec} shows the extinction opacities of the all three of these haze components by particle size.  At certain particle sizes, the opacities of these particles exhibit a slope significantly larger than that of the Rayleigh continuum, meaning that they could potentially be used to fit the observed rise, even in a hydrogen-poor atmosphere.  Since we find that hydrogen-rich models also fail to recreate this rise without a haze layer, we reach the conclusion that if the short-wavelength rise is valid, then GJ 1214b possesses a haze layer, which necessarily mutes molecular absorption features, but produces a significant rise in transit depth at short wavelengths.

For a haze layer to recreate the observed rise in transit depth, it must be thick enough to avoid being nearly transparent at all wavelengths, which would produce a negligible effect on the spectrum.  It must also be thin enough to avoid being nearly opaque at all wavelengths, since this would approach the limit of the opaque cloud layer, which also fails to agree with observations at short wavelengths.  These limits vary with the particle density and the thickness of the haze layer.

\begin{figure}[htp]			
\begin{center}
\includegraphics[scale=0.9]{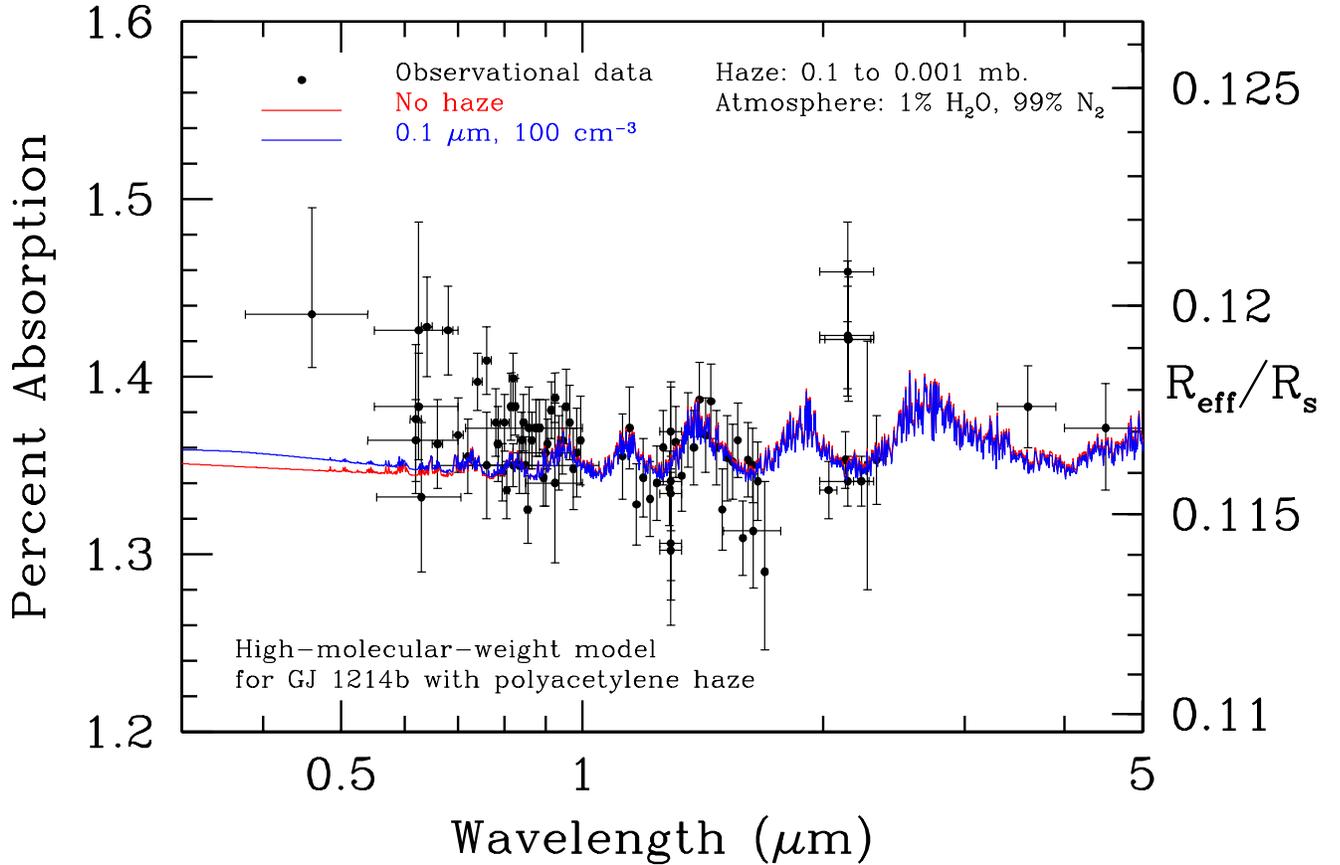} 
\end{center}
\caption{Transit depth in percent vs. wavelength in microns for a representative model for GJ 1214b with a 1\% ${\rm H_2O}$, 99\% ${\rm N_2}$ atmosphere with monodispersed polyacetylene haze.  The pressure range of the haze layer is chosen to be 0.1-0.001 mbar to maximize the contribution to extinction at large radii, above the principal pressure levels probed by Rayleigh scattering, thus increasing the slope of the Rayleigh scattering tail.  This haze model has a vertical optical depth of $\tau = 0.0034$ at 0.85 ${\rm \mu m}$, the midpoint of the normalization band.  The low slope of the Mie opacity curve (Figure \ref{haze-xsec}) results in a similarly low slope of the resulting spectrum, only slightly larger than the pure Rayleigh tail without haze.  There is very little effect on the rest of the spectrum.  The resulting slope remains too low to account for the observed tail, so, if the visible-wavelength data are valid, we do not support a model with a polyacetylene haze in a high-molecular-weight atmosphere.}
\label{poly-model}
\end{figure}

For polyacetylene particles of any size, the wavelength dependence is weak at short wavelengths.  A model representative of the effects of monodispersed polyacetylene haze in a hydrogen-poor atmosphere is shown in Figure \ref{poly-model}.  The pressure range of the haze layer is chosen to be 0.1-0.001 mbar to maximize the contribution above Rayleigh scattering to total extinction at large radii, which should increase the slope of the scattering tail.  This haze model has a vertical optical depth of $\tau = 0.0034$ at 0.85 ${\rm \mu m}$, the midpoint of the normalization band.  The small slope of the opacity curve results in a similarly small slope of the scattering spectrum, only slightly larger than the pure Rayleigh tail with no haze.  This has a very slight effect on the rest of the spectrum.  The resulting slope remains too small to account for the observed tail, so we do not support a polyacetylene haze in a high-molecular weight atmosphere as a model of GJ 1214b.

\begin{figure}[htp]			
\begin{center}
\includegraphics[scale=0.9]{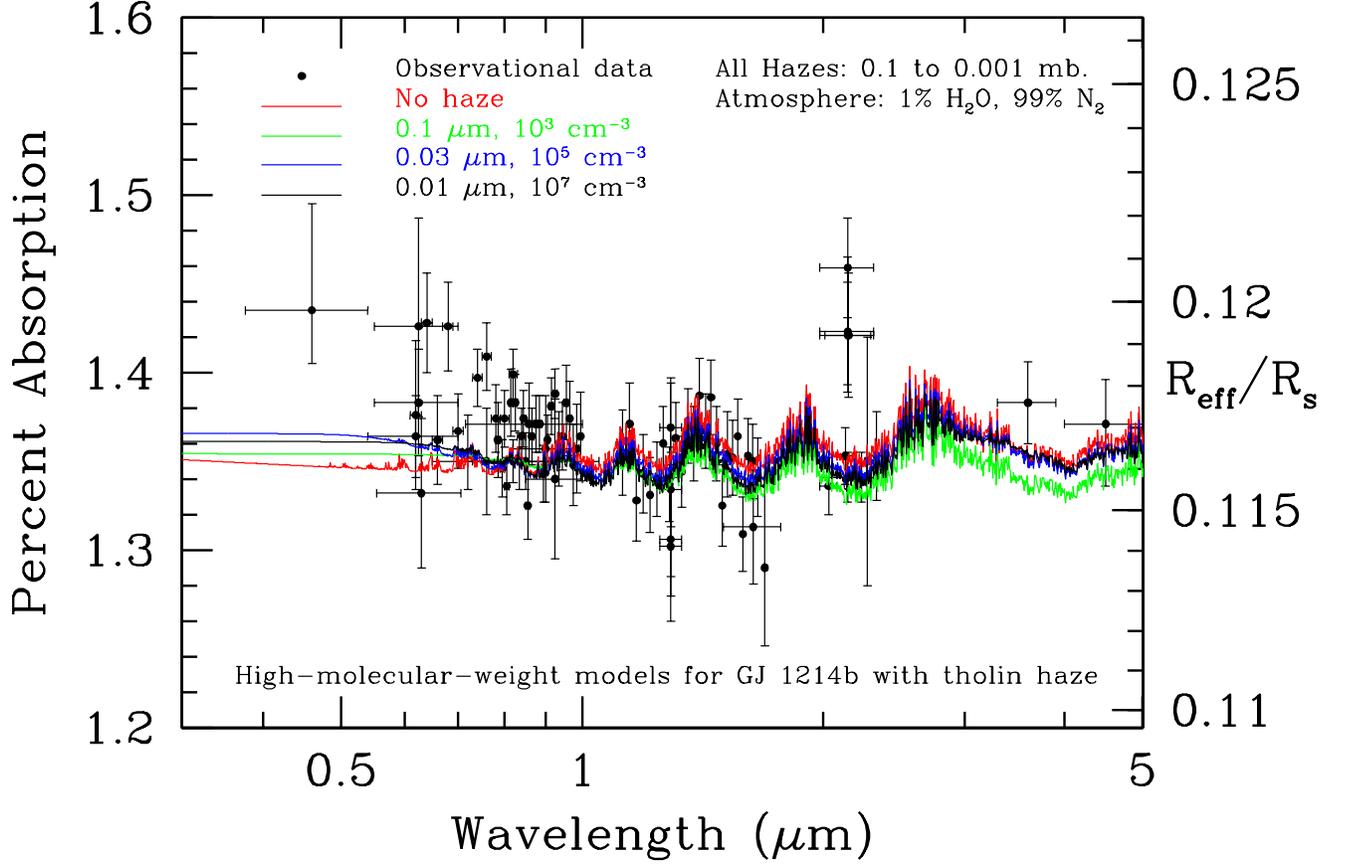} 
\end{center}
\caption{Transit depth in percent vs. wavelength in microns for atmosphere models of GJ 1214b with a 1\% ${\rm H_2O}$, 99\% ${\rm N_2}$ atmosphere with monodispersed tholin haze.  Each haze model has the same geometric depth.  The haze layer is taken to be between 0.1 mbar and 0.001 mbar.  Changes in transit depth in the 0.7-1.0 ${\rm \mu m}$ band affect the normalization, but all of the models are about equally consistent with the infrared observations.  At 0.85 ${\rm \mu m}$, the middle of this normalization band, the 0.1 ${\rm \mu m}$ haze model has a vertical optical depth of $\tau = 0.068$.  The corresponding values for the 0.03 ${\rm \mu m}$ and 0.01 ${\rm \mu m}$ haze models are $\tau = 0.010$ and 0.021, respectively.  At short wavelengths, the slope of the rise is greatest with the smallest particles, 0.01 ${\rm \mu m}$, but is still less than that of the observations.  This model is consistent with observations if the data points with the largest transit depths are excluded.}
\label{tholin-model}
\end{figure}

The largest slope for haze extinction opacities occurs with very small ($\lesssim \, 0.03 \, {\rm \mu m}$) tholin particles, making this the best candidate to fit the observed rise in a hydrogen-poor atmosphere.  Figure \ref{tholin-model} portrays several models with monodispersed tholin haze with the same geometric optical depth.  As before, the haze layer is put between 0.1 mbar and 0.001 mbar.  At 0.85 ${\rm \mu m}$, the middle of the normalization band, the vertical optical depth of the haze component in the 0.1 ${\rm \mu m}$ model is $\tau = 0.068$.  The corresponding values for the 0.03 ${\rm \mu m}$ and 0.01 ${\rm \mu m}$ haze models are $\tau = 0.010$ and 0.021, respectively.  Changes in transit depth in the 0.7-1.0 ${\rm \mu m}$ band affect the normalization, but all of the models are about equally consistent with the infrared observations.  At short wavelengths, the slope of the rise in transit depth is greatest with the smallest particles, 0.01 ${\rm \mu m}$.  This is consistent with observations if the data points with the largest transit depths are excluded.  There is some evidence for this idea in the qualitative bifurcation of the data blueward of 0.8 ${\rm \mu m}$, although there are measured points from both Bean et al. (2011) and from de Mooij et al. (2011) at both large and small transit depths.

The problem of hydrogen-rich models with hazes presents similar constraints.  The slope required to reproduce the observed short-wavelength rise is a less strict constraint because of the lower molecular weight of the atmosphere, a fact that justifies a renewed interest in polyacetylene haze.  However, a hydrogen-rich atmosphere provides an additional constraint in that large opacities are needed in the near- and mid-infrared to mute the large absorption features, especially the suggested 1.5 ${\rm \mu m}$ feature.

\begin{figure}[htp]			
\begin{center}
\includegraphics[scale=0.9]{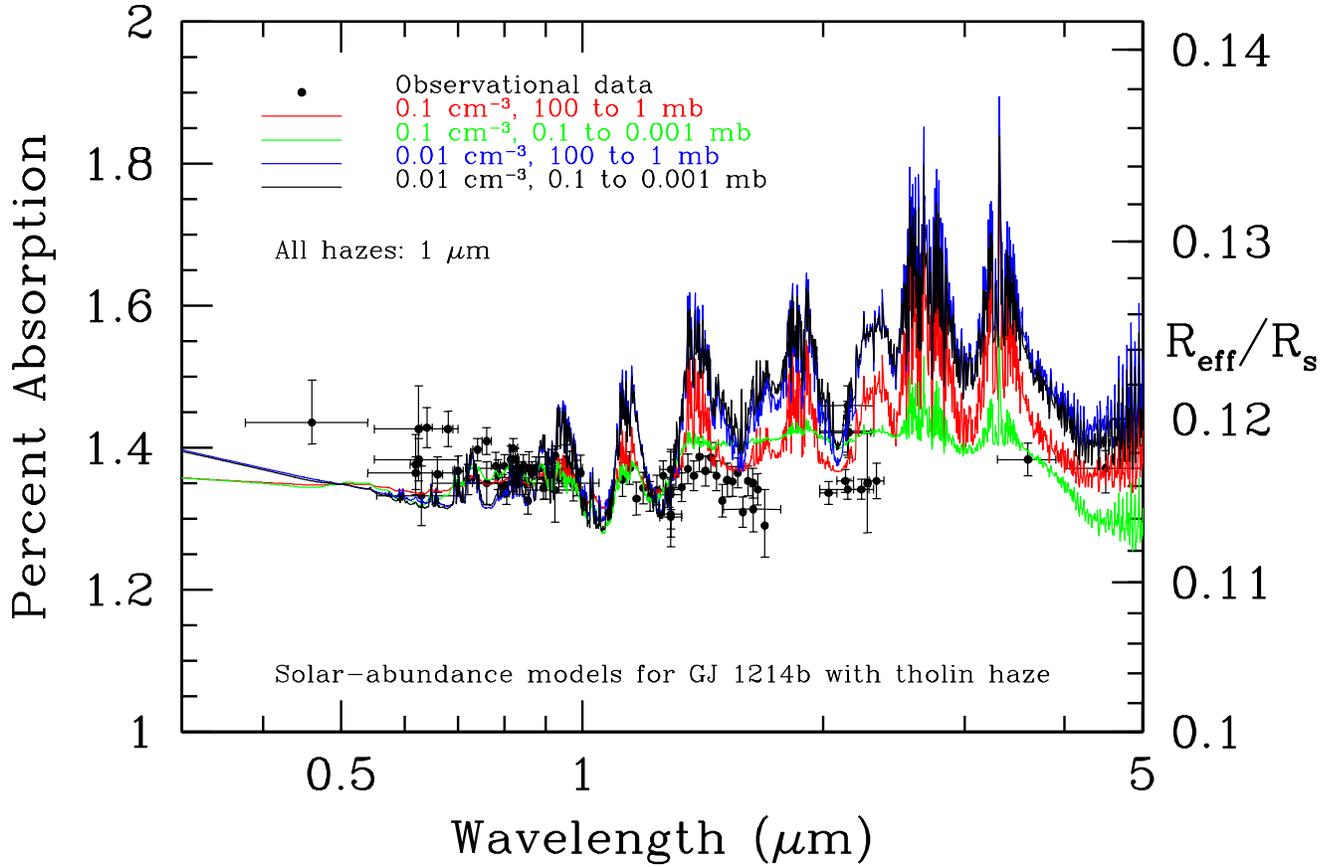} 
\end{center}
\caption{Transit depth in percent vs. wavelength in microns for atmospheric models of GJ 1214b with a solar-abundance atmosphere with monodispersed tholin haze with a particle size of 1.0 ${\rm \mu m}$ at various pressure levels.  At 0.85 ${\rm \mu m}$, the vertical optical depths of the models are 0.20 for a particle density of $0.1 \, {\rm cm^{-3}}$ and 0.020 for a particle density of $0.01 \, {\rm cm^{-3}}$.  Because the wavelength dependence of the haze opacity is relatively flat, the amplitudes of the spectral features are uniformly suppressed.  Thus, the slope of the short-wavelength tail is low and inconsistent with the data.  For some choices of haze altitude and particle density, the near- and mid-infrared features begin to conform with the data.}
\label{tholin1}
\end{figure}

From Figure \ref{haze-xsec}, the wavelength dependence of both haze species we consider is very similar at particle sizes of 1 ${\rm \mu m}$ and larger.  Furthermore, the wavelength dependence is almost completely flat for 10 ${\rm \mu m}$ particles and is relatively flat with 1 ${\rm \mu m}$ particles, since the particles are similar in size or larger than the wavelengths of light of interest, so their effects are roughly proportional to their geometric cross sections.  Figure \ref{tholin1} shows the effect of a solar-abundance atmosphere with various monodispersed tholin haze models with a particle size of 1 ${\rm \mu m}$.  At 0.85 ${\rm \mu m}$, the vertical optical depths of the models are 0.20 for a particle density of $0.1 \, {\rm cm^{-3}}$ and 0.020 for a particle density of $0.01 \, {\rm cm^{-3}}$.  Because of the flat wavelength dependence, especially in the visible, the resulting spectra are flattened uniformly to varying degrees, so that the short-wavelength rise is suppressed, and the model fails to achieve the goal of fitting the short-wavelength data.  However, with a relatively large particle density of $0.1 \, {\rm cm^{-3}}$ and a high altitude of 0.1-0.001 mbar, the spectrum begins to conform with the near- and mid-infrared data.  While it is easier to keep small particles aloft, we do not know the degree of turbulence in GJ 1214b's atmosphere, so we do not make any judgment concerning the plausibility of these conditions.  Further research is needed to investigate this issue.

\begin{figure}[htp]			
\begin{center}
\includegraphics[scale=0.9]{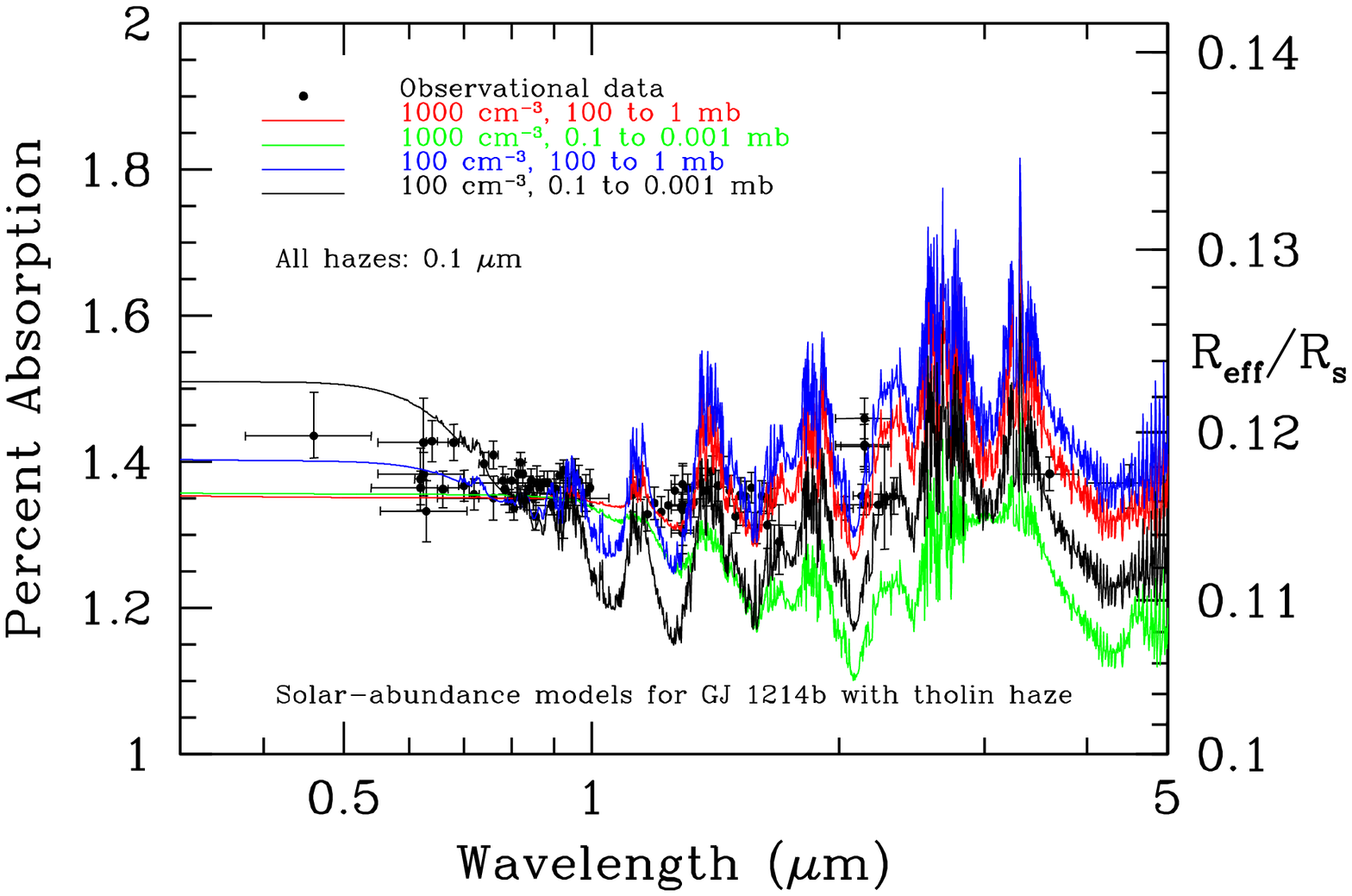} 
\end{center}
\caption{Transit depth in percent vs. wavelength in microns for atmospheric models of GJ 1214b with a solar-abundance atmosphere with monodispersed tholin haze with particle size a 0.1 ${\rm \mu m}$ at various pressure levels.  At 0.85 ${\rm \mu m}$, the vertical optical depths of the models are 0.82 for a particle density of $1000 \, {\rm cm^{-3}}$ and 0.082 for a particle density of $100 \, {\rm cm^{-3}}$.  These spectra can fit the observed short-wavelength rise in the complete data set, as well as subsets including high-depth data points, low-depth data points, and the subset of data points from each source, depending on the pressure level of the haze layer.  However, the lower opacities in the near- and mid-infrared result in corresponding infrared features that are too large in amplitude to fit the data.}
\label{tholin0.1}
\end{figure}

Figure \ref{tholin0.1} shows a set of models with hydrogen-rich atmospheres with monodispersed tholin hazes with a particle size of 0.1 ${\rm \mu m}$.  At 0.85 ${\rm \mu m}$, the vertical optical depths of the models are 0.82 for a particle density of $1000 \, {\rm cm^{-3}}$ and 0.082 for a particle density of $100 \, {\rm cm^{-3}}$.  This figure demonstrates a problem that occurs in a large portion of the parameter space of hazes.  While the haze dominates the atmospheric opacity at visible wavelengths, its opacity becomes too small at infrared wavelengths, resulting in infrared features that remain too large in amplitude to fit the data.  This problem often persists even if the haze opacity saturates in the visible, and the layer becomes uniformly opaque at short wavelengths.  Depending upon the choice of parameters, these models can be made to fit the short-wavelength rise in the data and could do so even if some of these data were refuted.  However, in each case, the infrared features either are far too large to fit the data or are normalized to an incorrect transit depth.

\begin{figure}[htp]			
\begin{center}
\includegraphics[scale=0.9]{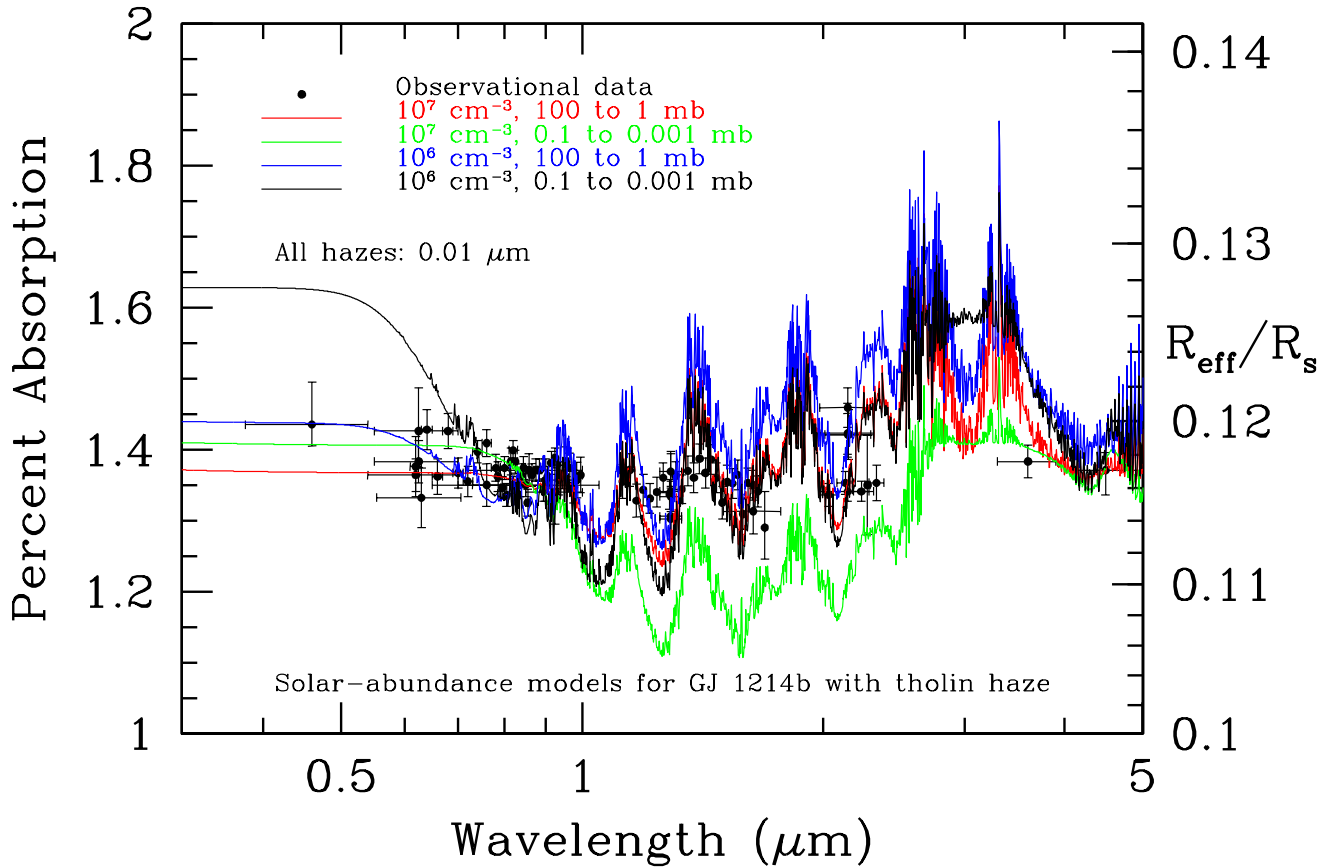} 
\end{center}
\caption{Transit depth in percent vs. wavelength in microns for atmospheric models of GJ 1214b with a solar-abundance atmosphere with monodispersed tholin haze with a particle size of 0.01 ${\rm \mu m}$ at high altitudes (between the 0.1 and 0.001 mbar levels).  At 0.85 ${\rm \mu m}$, the vertical optical depths of the models are 0.26 for a particle density of $10^7 \, {\rm cm^{-3}}$ and 0.026 for a particle density of $10^6 \, {\rm cm^{-3}}$.  These spectra predict a short-wavelength rise that is steeper than in the observed data, such that the transit depth saturates near 0.5 ${\rm \mu m}$.  Any increase in particle density would increase the amount of saturation and result in a poorer fit to the data.  Still, the lower opacity in the infrared results in infrared features that are too large in amplitude to fit the data.  A decrease in haze altitude would also make the molecular features more prominent.  Therefore, tholin particles in this size range in a hydrogen-rich atmosphere can be ruled out as an explanation of the spectrum of GJ 1214b.}
\label{tholin0.01}
\end{figure}

Figure \ref{tholin0.01} provides models with hydrogen-rich atmospheres with monodispersed tholin hazes with a particle size of 0.01 ${\rm \mu m}$.  At 0.85 ${\rm \mu m}$, the vertical optical depths of the models are 0.26 for a particle density of $10^7 \, {\rm cm^{-3}}$ and 0.026 for a particle density of $10^6 \, {\rm cm^{-3}}$.  The results are much the same as for the 0.1 ${\rm \mu m}$ models.  The opacity saturates at short wavelengths, sometimes resulting in a flat spectrum in the visible, while lower infrared opacities result in near- and mid-infrared features that are too large to fit the data.  In other cases, the opacity does not saturate at these wavelengths, but the slope of the predicted short-wavelength tail is much larger than is observed.  Because of the difficultly of fitting the short-wavelength rise and the improbability of particles this small, we do not advance this model as an explanation for the observed features.

Polyacetylene has a spectral behavior similar to tholins at larger particle sizes ($\gtrsim$1 ${\rm \mu m}$), and at smaller particle sizes ($\lesssim$0.1 ${\rm \mu m}$), it suffers from even lower infrared opacities than tholins, so the same problems persist that occur for tholins.  A haze model that might successfully fit both the observed short-wavelength rise in the data and the low amplitudes of the infrared features is one for which the opacity follows a Rayleigh-like $\lambda^{-4}$ behavior blueward of 1 ${\rm \mu m}$ and is roughly constant redward of 1 ${\rm \mu m}$.  None of the hazes we test fits this description.

\subsection{Best-fit Models}
\label{best}

\begin{figure}[htp]			
\begin{center}
\includegraphics[scale=0.9]{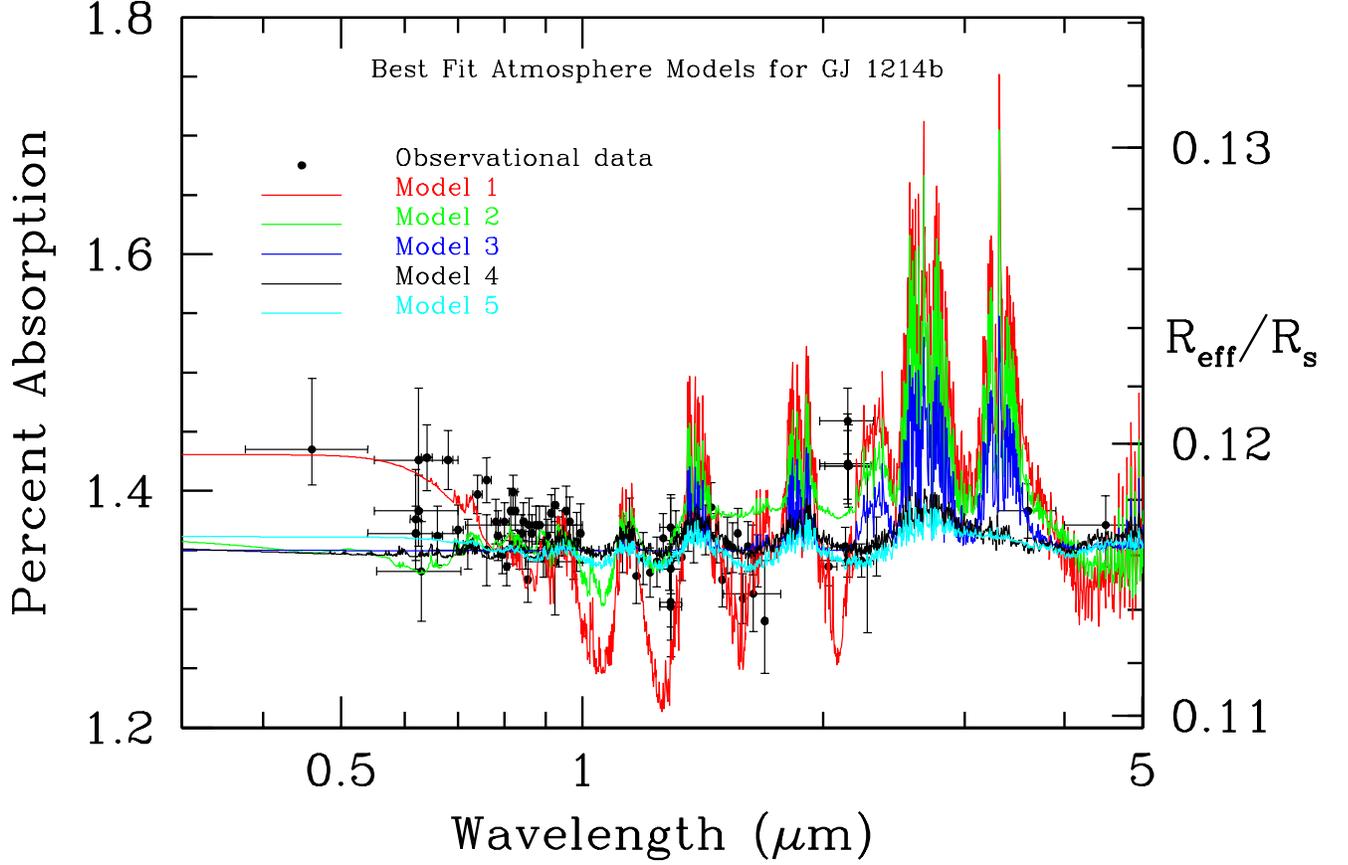} 
\end{center}
\caption{Transit depth in percent vs. wavelength in microns for our best fit atmosphere models of GJ 1214b.  Models 1-3 use a solar-abundance atmosphere, while Models 4 and 5 use an atmosphere of 1\% ${\rm H_2O}$ and 99\% ${\rm N_2}$.  In Model 1, our best fit to the short-wavelength rise in the data, there is a tholin haze layer (although the choice of haze species is arbitrary for Models 1 and 2) with a particle size of 0.1 ${\rm \mu m}$, a particle density of 100 ${\rm cm^{-3}}$, and a pressure range of 10-0.1 mbar (vertical optical depth at 0.85 ${\rm \mu m}$: 0.082).  Model 2 uses a tholin haze layer with a particle size of 1 ${\rm \mu m}$, a particle density of 0.1 ${\rm cm^{-3}}$, and a pressure range of 10-0.1 mbar (vertical optical depth at 0.85 ${\rm \mu m}$: 0.20).  Model 3 includes a uniformly opaque cloud layer with cloud tops at 0.3 mbar.  Models 2-4 are our best fits if larger transit depths are excluded.  Model 4 has no clouds or haze, but is not very sensitive to the addition of any haze that does not significantly obscure the absorption features, or to the proportion of water (within an order of magnitude).  Model 5 is our best fit to the short-wavelength rise with a high-molecular-weight atmosphere.  It includes tholin haze with a particle size of 0.01 ${\rm \mu m}$, a particle density of $10^6 \, {\rm cm^{-3}}$, and a pressure range of 0.1-0.001 mbar (vertical optical depth at 0.85 ${\rm \mu m}$: 0.021).  However, it is consistent with only the lower edge of the short-wavelength data distribution and is inconsistent with many of the data points in the distribution.  Therefore, we conclude that if this rise is valid, GJ 1214b should have a hydrogen-rich atmosphere.}
\label{1214b-best}
\end{figure}

Our best fit models for GJ 1214b are given in Figure \ref{1214b-best}.  We select five models by visual inspection.  While this is not, in principle, the most accurate method, the systematic problems with the data set indicate that any quantitative assessment of the fits would not necessarily be any better.  Moreover, many of our theoretical spectra are similar enough that a range of parameters of the haze layer (in some cases as large as a factor of two) would produce very similar fits.  Thus, any quantitative fit would not produce a significant gain in precision over our "eyeball" technique.

Of our five selections, Model 1 is the only model we select that provides a good fit to the observed short-wavelength rise in transit depth.  It utilizes a solar-abundance atmosphere with a tholin haze (although the choice of haze species is arbitrary) with a particle size of 0.1 ${\rm \mu m}$, a particle density of 100 ${\rm cm^{-3}}$, and a pressure range of 10 to 0.1 mbar.  This model has a vertical optical depth at 0.85 ${\rm \mu m}$ of $\tau = 0.082$.  It is consistent with many of the near- and mid- infrared data points, but the predicted spectral features have much larger amplitudes than are observed.

Models 2-4 are our best fits to the data points with low transit depths, with larger transit depths excluded.  Model 2 does this with a solar-abundance atmosphere and a translucent tholin haze layer (the choice of haze species being arbitrary) with a particle size of 1 ${\rm \mu m}$, a particle density of 0.1 ${\rm cm^{-3}}$, and residing in a pressure range from 10-0.1 mbar; it has a vertical optical depth at 0.85 ${\rm \mu m}$ of $\tau = 0.20$.  Model 3 instead uses solar abundances with a uniformly opaque cloud layer, with cloud tops at 0.3 mbar.  This model provides a better fit in the near- and mid-infrared than the translucent haze.

Model 4 is our overall best fit using a high-molecular-weight atmosphere, specifically, 1\% ${\rm H_2O}$ and 99\% ${\rm N_2}$ with no clouds or haze.  However, this model is not very sensitive to the proportion of water (within an order of magnitude) or to the addition of any haze that does not significantly obscure the absorption features.

Model 5 is our best fit to the short-wavelength rise with a high-molecular-weight atmosphere.  It includes a 1\% ${\rm H_2O}$ and 99\% ${\rm N_2}$ atmosphere with a tholin haze with a particle size of 0.01 ${\rm \mu m}$, a particle density of $10^6 \, {\rm cm^{-3}}$, and residing in a pressure range from 0.1 to 0.001 mbar; it has a vertical optical depth at 0.85 ${\rm \mu m}$ of $\tau = 0.021$.  Here, the choice of tholins is not arbitrary.  Polyacetylene has a near-discontinuous wavelength dependence in the near-infrared at this particle size, which makes it unsuitable for this model.  This haze model is relatively improbable because of its small particle size, and it is consistent only with the lower edge of the observed short-wavelength rise in transit depth.  Therefore, we conclude that if this rise is valid, GJ 1214b should have a hydrogen-rich atmosphere.

A hydrogen-rich model provides a better numerical fit to the large-transit-depth data, and the systematics predict a low mean molecular weight for the atmosphere if the short-wavelength rise is valid.  Furthermore, while it is possible to fit the low amplitudes of the observed near- and mid-infrared features with a hydrogen-rich model, it is not possible to fit the large-amplitude features with a hydrogen-poor model.  Therefore, we conclude that a hydrogen-rich atmosphere is more probable overall for GJ 1214b, although we cannot rule out a hydrogen-poor model based on the current data.  Because a hydrogen-rich model predicts larger-amplitude features, observations of the transit depth around 2.7 and 3.3 ${\rm \mu m}$, and to a lesser extent near 1.9 ${\rm \mu m}$, could be valuable for distinguishing hydrogen-rich from hydrogen-poor models.

We are unable to produce a model that fits all of the data, even after omitting either the ${\rm K_s}$-band or K-band measurements.  Indeed, the best fit overall is the unacceptably improbable 0.01$\times$ solar model with no haze.  If all of the non-conflicting data are valid, then we can say with confidence that there is a haze layer, but we do not make any predictions about its properties.  Additionally, our methodology rules out any hazeless hydrogen-rich model in general and hydrogen-rich models with tholin hazes in the 0.01 ${\rm \mu m}$ range, as well as any model in which the largest source of absorption is not water.  We look forward to refining our model in light of additional data.




\section{Discussion and Conclusions}
\label{conclusion}

We have developed a new capability to compute transit spectra of super-Earths that allows atmospheres with arbitrary proportions of common molecular species, along with hazes.  Similar hazes have been used to characterize the atmospheres of giant planets, and they appear to be important in explaining the apparently conflicting observations of GJ 1214b.  However, additional measurements are needed in a range of wavelengths to determine this object's transit spectrum more precisely.

If the observed short-wavelength rise is valid, then GJ 1214b most likely possesses a hydrogen-rich atmosphere with a haze of small ($\sim$0.1 ${\rm \mu m}$) particles, although there is also the possibility of an atmosphere rich in nitrogen and water, with very small ($\sim$0.01 ${\rm \mu m}$), very high altitude (up to 0.001 mbar) tholin haze particles.  The latter model is a poorer fit in the visible and is physically less probable, but it cannot be ruled out by the current data.  However, the hydrogen-rich model provides a very poor fit redward of 1 ${\rm \mu m}$.

Alternatively, the short-wavelength measurements may prove inaccurate, but the near-infrared measurements may be accurate, including the 1.5-${\rm \mu m}$ suspected water feature and the low-amplitude $K$-band measurements.  This is a particularly plausible scenario because of the greater stellar variability and lower luminosity for M-dwarfs in the optical.  In this case, in order of goodness of fit, GJ 1214b likely possesses (1) an ${\rm N_2-H_2O}$ atmosphere, (2) a solar-abundance atmosphere, with thick, opaque clouds at or above the 1 mbar level, or (3) a solar-abundance atmosphere with a translucent haze with medium-to-large ($\gtrsim$1 ${\rm \mu m}$) particles.  For hydrogen-rich atmospheres, the models are not sensitive to the haze composition.  High-molecular-weight models are not sensitive to the addition of hazes or clouds in general or to the abundance of water, within an order of magnitude.  None of these models shows a short-wavelength rise in transit depth or a $K_s$-band feature as deep as the one suggested by the data.

In both cases, our calculations rule out a hazeless (and cloudless) hydrogen-rich atmosphere, which predicts a transit depth in the optical that is much smaller than the data for any reasonable metallicity.  We also rule out a hydrogen-rich atmosphere with tholin particles in the $\sim$0.01 ${\rm \mu m}$ range and a high-molecular-weight atmosphere with significant amounts of absorbing species other than water, such as ${\rm NH_3, CO, or \, CO_2}$, which is inconsistent with the near-infrared data.  Overall, the prospects for fitting the data appear better with a hydrogen-rich atmosphere, but we cannot rule out a hydrogen-poor composition.

Some of the data for GJ 1214b conflict, particularly those in the $K$ band and $K_s$ band.  Even for the data that do not directly conflict, none of the models we test provides a good fit to all of the observed features simultaneously.  However, we can rule out large classes of models that do not fit any of the observed features.  In the case that all of the non-conflicting data are valid, we do not support any model without a haze layer.

We propose additional observational tests to distinguish hydrogen-rich from hydrogen-poor atmospheres.  All of our hydrogen-rich models predict a strong water feature at 2.7 ${\rm \mu m}$ and a strong methane feature at 3.3 ${\rm \mu m}$, both of which should be unambiguously larger in amplitude than the features predicted by any reasonable hydrogen-poor model.  The hydrogen-rich models also predict a water feature at 1.9 ${\rm \mu m}$ that could give an unambiguous result depending on the actual atmospheric parameters.  Therefore, data at these wavelengths would provide strong evidence to distinguish among these models.

Our models correctly reproduce the expected systematics of transit spectra of planets with varying metallicities, temperatures, and cloud layers.  They also reproduce the expected features from molecular absorption.  Nevertheless, our models use an isothermal {\it T-P} profile.  While this is a good first-order approximation, a more accurate {\it T-P} profile is desired to produce a more accurate transit spectrum, along with an atmospheric circulation model of the behavior at the terminator, which will not be in thermal or chemical equilibrium.  A photochemical model is also desired to determine the abundances of molecular species created or destroyed by photolysis.

We have prepared an online library of theoretical transit spectra produced by our calculations, sampling the known and expected parameter space of super-Earths.  This library includes models with equilibrium temperatures from 300 K to 1000 K and silicate-iron planets from 1 ${M_\Earth}$ to 10 ${M_\Earth}$.  The library also includes models with atmospheric compositions of solar abundance, 0.3$\times$ solar, 3$\times$ solar, and pure ${\rm H_2O, CO_2, and \, CO}$ atmospheres, as well as a variety of haze models, including sulfuric acid hazes.

Very recently, we learned of the work of Murgas et al. (2012), who measured the transit depth of GJ 1214b at three additional wavelengths in the $R$ band.  For completeness, these data points are included in Figure \ref{Data}.  These points are consistent with the proposed rise in transit depth at short wavelengths; however, their uncertainties are too large to make any definite conclusions.

\acknowledgments

We thank Matteo Brogi and Zachory Berta for providing data in electronic form prior to publication, and Jacob Bean, Ian Crossfield, and Jean-Michel D\'{e}sert for helpful conversations, comments, and suggestions.  The authors also acknowledge support in part under NASA ATP grant NNX07AG80G, {\it HST} grants HST-GO-12181.04-A, HST-GO-12314.03-A, and HST-GO-12550.02, and JPL/Spitzer Agreements 1417122, 1348668, 1371432, 1377197, and 1439064.  We have prepared a large suite of models of transit spectra of super-Earths for a range of masses, radii, temperatures, and compositions.  These models are made available at http://www.astro.princeton.edu/$\sim$arhowe and \\ http://www.astro.princeton.edu/$\sim$burrows.

{}
\clearpage

\begin{appendix}
\section{Analytic Model}
\label{analytic}

Lecavelier des Etangs et al. (2008) presented an analytic model of Rayleigh scattering in a planetary atmosphere as seen in transit.  Here, we investigate the circumstances under which their model can also be applied to absorption.  We find that the model accurately predicts the results of numerical calculations if the absorption opacity of the atmosphere is low, i.e., of the same order as the Rayleigh scattering opacity, but it significantly underpredicts the transit depth if the opacity is high.  The fit is good enough at low opacities to justify this model as an estimator of the mean molecular weight of the atmosphere of GJ 1214b based on the slope of the short-wavelength rise in the data.

The analytic model begins with an approximation of the optical depth of transmission at an impact parameter $z$,
\begin{equation}
\tau(\lambda,z) \approx \sigma(\lambda)n(z)\sqrt{2\pi R_pH},
\end{equation}
where $R_p$ is the planetary radius, $H$ is the scale height, and $n(z) = n_0{\rm exp}(-z/H)$ is the particle density ($n_0$ being the density at the surface or reference altitude).  Then, if $R_{\rm obs}$ is the observed radius of the planet at a particular wavelength, one can define $\tau_{\rm eq}$ by $R_{\rm obs} = R_p+z(\tau_{\rm eq})$.  That is, $\tau_{\rm eq}$ is the optical depth at the impact parameter of the planet's originally observed radius.  Numerical calculations suggest that for a wide range of scale heights, $\tau_{\rm eq} \approx 0.56$.  Thus, for a given wavelength of transmitted light, 
\begin{equation}
0.56 \approx \sigma(\lambda)n(z(\lambda))\sqrt{2\pi R_{\rm obs}(\lambda)H},
\end{equation}
or, solving for $z$, the planet's effective radius lies at an altitude of,
\begin{equation}
z(\lambda) = H \ln(P_0 \sigma_{\rm avg}(\lambda)/0.56 \times \sqrt{2\pi R_{\rm obs}/kT\mu g}),
\end{equation}
where $\sigma_{\rm avg}$ is the average cross section per particle and $P_0$ is the base pressure.

The analytic model is valuable because it yields a single relation between the effective radius, temperature, and mean molecular weight of the atmosphere.  If the wavelength dependence of the cross section is known, as it is for Rayleigh scattering, this formula allows a derivation of the temperature of the planet from its spectrum, since
\begin{equation}
\frac{dz}{d\lambda} \approx H\frac{d\ln{\sigma}}{d\lambda}.
\end{equation}
If $\sigma_{\rm tot}=(\lambda/\lambda_0)^\alpha$, then,
\begin{equation}
\frac{dR_{\rm obs}}{d\ln{\lambda}} = \frac{dz}{d\ln{\lambda}} = \alpha H,
\end{equation}
or, given $H = kT/\mu g$,
\begin{equation}
\alpha T = \frac{\mu g}{k} \frac{dR_{\rm obs}}{d\ln \lambda}.
\end{equation}
Alternatively, if the temperature is known, we can solve for the mean molecular weight:
\begin{equation}
\mu = \frac{\alpha T k}{g} \frac{d\ln \lambda}{dR_{\rm obs}}.
\end{equation}
Plugging in the parameters for GJ 1214b yields a mean molecular weight of about 2.7.

This model assumes that opacity is independent of temperature and pressure, which is nearly correct for Rayleigh scattering, but a poor approximation for molecular absorption.  In the Rayleigh scattering regime, at short wavelengths where it dominates over absorption, we find that this model predicts the pressure at the effective radius within a few percent of numerical calculations.  This corresponds to much less than one scale height, so the resultant transit depth is within a fraction of a percent of the numerical value.  For absorption, the agreement is still within a factor of a few at most wavelengths, but the numerical calculations have a larger dispersion in effective radius.  For wavelengths where the opacity is very high, this discrepancy can reach a factor of 100 in pressure, or about 5 scale heights.

The dispersion and the discrepancy at the highest opacities are likely due to the analytic model's approximation that the opacity of the atmosphere at each wavelength is constant.  Molecular absorption is often roughly constant over a wide pressure range, but it varies significantly at high pressures deep in the atmosphere and at the low pressures probed by very high opacities.  Thus, the analytic model does a good job for Rayleigh scattering and when the total opacity is relatively low, probing higher pressures, but it becomes less accurate in cases where opacities are higher or are highly variable with temperature and pressure.  However, it can be made accurate for particular pressure levels by a judicious choice of the ``average'' opacity.
\end{appendix}

\begin{center}
\begin{deluxetable}{ccccccc}

\tablewidth{475pt}

\tablecaption{Properties of Transiting Earths and Super-Earths}
\tablehead{
Name & Radius     & Mass       & $a$ & Period & $i$ & $e$ \\
     & $(R_\Earth)$ & $(M_\Earth)$ & (AU) & (d) & (deg) &     }
\startdata
GJ 1214b    & $2.678\pm 0.13$ & $6.55\pm 0.98$ & 0.0143                    & 1.5803925 & $88.62^{+0.36}_{-0.28}$    & $<0.27$      \\
55 Cancri e\tablenotemark{1} & $2.00\pm 0.14$  & $8.63\pm 0.35$ & 0.01560                   & 0.736537  & 83.4                       & $0.17\pm 0.04$ \\
COROT-7b\tablenotemark{2}    & $1.58\pm 0.1$   & $<21$\tablenotemark{3} & 0.0172            & 0.85358   & $80.1\pm 0.3$              & ~0           \\
Kepler-10b\tablenotemark{4}  & $1.416^{+0.033}_{-0.036}$ & $4.56^{+1.17}_{-1.29}$ & 0.01684 & 0.837495  & 84.4                       & ~0           \\
Kepler-22b\tablenotemark{5}  & $2.38\pm 0.13$  & $<82$\tablenotemark{3} & 0.849             & 289.8623  & $89.764^{+0.025}_{-0.042}$ &              \\
Venus\tablenotemark{6}       & $0.950$         & $0.815$        & 0.728                     & 224.7     & 86.61                      & 0.0068       \\
\enddata
\label{planets}
\tablenotetext{1}{From Demory et al. (2011).}
\tablenotetext{2}{From L\'{e}ger et al. (2009).}
\tablenotetext{3}{2$\sigma$ limit.}
\tablenotetext{4}{From Batalha et al. (2011).}
\tablenotetext{5}{From Borucki et al. (2011).}
\tablenotetext{6}{Basilevsky \& Head (2003).}
\end{deluxetable}
\end{center}

\begin{center}
\begin{deluxetable}{cccccccc}
\tablewidth{450pt}
\tablecaption{Properties of Host Stars of Transiting Super-Earths\tablenotemark{1}}
\tablehead{
Name & $m_V$ & Radius    & Mass      & $T_{eff}$ & Luminosity & [Fe/H] & log(g)      \\
     &       & $(R_\odot)$ & $(M_\odot)$ & (K)   & $(L_\odot)$ &       & ${\rm (cm s^{-2})}$ }
\startdata
GJ 1214    & 14.67     & 0.2110 & 0.157 & 3026 & 0.00328 & +0.39\tablenotemark{2} & 4.991 \\
55 Cancri\tablenotemark{3}  & 5.95      & 0.943  & 0.963 & 5234 & 0.582\tablenotemark{4} & +0.31   & 4.45  \\
COROT-7\tablenotemark{5}    & 11.668    & 0.87   & 0.93  & 5275 & 0.49\tablenotemark{6} & +0.03\tablenotemark{7} & 4.50  \\
Kepler-10\tablenotemark{8}  & $\sim 11$ & 1.051  & 0.895 & 5627 & 1.004   & -0.15                  & 4.35  \\
Kepler-22\tablenotemark{9}  & 11.664    & 0.979  & 0.970 & 5518 & 0.79    & -0.29                  & 4.44  \\
\enddata
\label{stars}
\tablenotetext{1}{Uncertainties have been omitted for compactness.}
\tablenotetext{2}{From Berta et al. (2011a).}
\tablenotetext{3}{From Demory et al. (2011).}
\tablenotetext{4}{From von Braun et al. (2011).}
\tablenotetext{5}{From L\'{e}ger et al. (2009).}
\tablenotetext{6}{From Guenther et al. (2010).}
\tablenotetext{7}{[M/H].}
\tablenotetext{8}{From Batalha et al. (2011).}
\tablenotetext{9}{From Borucki et al. (2011).}
\end{deluxetable}
\end{center}

\begin{center}
\begin{deluxetable}{cccccccc}
\tablewidth{450pt}
\tablecaption{List of Atmosphere Models for GJ 1214b}
\tablehead{
Name       & Composition                  & Haze Type     & Haze Size    & Haze Density & Haze Level & Figure                   \\
           &                              &               & (${\rm \mu}$m) & (cm$^{-3}$) & (mb)      &                          }
\startdata
Solar      & 1 $\times$ solar abundances  & None          &              &              &            & \ref{solar-model}        \\
0.1Solar   & 0.1 $\times$ solar abundances & None          &              &              &            & \ref{metals-model}       \\
0.01 Solar & 0.01 $\times$ solar abundances & None          &              &              &            & \ref{metals-model}       \\
100mb      & Solar                        & Opaque clouds &              &              & 100        & \ref{clouds-model}       \\
10mb       & Solar                        & Opaque clouds &              &              & 10         & \ref{clouds-model}       \\
1mb        & Solar                        & Opaque clouds &              &              & 1          & \ref{clouds-model}       \\
0.1mb      & Solar                        & Opaque clouds &              &              & 0.1        & \ref{clouds-model}       \\
100h2o     & 100\% H$_2$O                 & None          &              &              &            & \ref{water-model}        \\
10h2o      & 10\% H$_2$O, 90\% N$_2$      & None          &              &              &            & \ref{water-model}        \\
Model 3    & 1\% H$_2$O, 99\% N$_2$       & None          &              &              &            & \ref{water-model},\ref{poly-model},\ref{tholin-model},\ref{1214b-best} \\
100mbh2o   & 100\% H$_2$O                 & Opaque clouds &              &              & 100        & \ref{water-clouds-model} \\
10mbh2o    & 100\% H$_2$O                 & Opaque clouds &              &              & 10         & \ref{water-clouds-model} \\
1mbh2o     & 100\% H$_2$O                 & Opaque clouds &              &              & 1          & \ref{water-clouds-model} \\
0.1mbh2o   & 100\% H$_2$O                 & Opaque clouds &              &              & 0.1        & \ref{water-clouds-model} \\
100ch4     & 100\% CH$_4$                 & None          &              &              &            & \ref{ch4}                \\
10ch4      & 10\% CH$_4$, 90\% N$_2$      & None          &              &              &            & \ref{ch4}                \\
1ch4       & 1\% CH$_4$, 99\% N$_2$       & None          &              &              &            & \ref{ch4}                \\
poly       & 1\% H$_2$O, 99\% N$_2$       & Polyacetylene & 0.1          & 100          & 0.1-0.001  & \ref{poly-model}         \\
tholin0.1  & 1\% H$_2$O, 99\% N$_2$       & Tholin        & 0.1          & 10$^3$       & 0.1-0.001  & \ref{tholin-model}       \\
tholin0.03 & 1\% H$_2$O, 99\% N$_2$       & Tholin        & 0.03         & 10$^5$       & 0.1-0.001  & \ref{tholin-model}       \\
tholin0.01 & 1\% H$_2$O, 99\% N$_2$       & Tholin        & 0.01         & 10$^7$       & 0.1-0.001  & \ref{tholin-model}       \\
solar1     & Solar                        & Tholin        & 1.0          & 0.1          & 100-1      & \ref{tholin1}            \\
solar2     & Solar                        & Tholin        & 1.0          & 0.1          & 0.1-0.001  & \ref{tholin1}            \\
solar3     & Solar                        & Tholin        & 1.0          & 0.01         & 100-1      & \ref{tholin1}            \\
solar4     & Solar                        & Tholin        & 1.0          & 0.01         & 0.1-0.001  & \ref{tholin1}            \\
solar5     & Solar                        & Tholin        & 0.1          & 1000         & 100-1      & \ref{tholin0.1}          \\
solar6     & Solar                        & Tholin        & 0.1          & 1000         & 0.1-0.001  & \ref{tholin0.1}          \\
solar7     & Solar                        & Tholin        & 0.1          & 100          & 100-1      & \ref{tholin0.1}          \\
solar8     & Solar                        & Tholin        & 0.1          & 100          & 0.1-0.001  & \ref{tholin0.1}          \\
solar9     & Solar                        & Tholin        & 0.01         & 10$^7$       & 100-1      & \ref{tholin0.01}         \\
solar10    & Solar                        & Tholin        & 0.01         & 10$^7$       & 0.1-0.001  & \ref{tholin0.01}         \\
solar11    & Solar                        & Tholin        & 0.01         & 10$^6$       & 100-1      & \ref{tholin0.01}         \\
Model 5    & Solar                        & Tholin        & 0.01         & 10$^6$       & 0.1-0.001  & \ref{tholin0.01},\ref{1214b-best} \\
Model 1    & Solar                        & Tholin        & 0.1          & 100          & 10-0.1     & \ref{1214b-best}         \\
Model 2    & Solar                        & Tholin        & 1.0          & 0.1          & 10-0.1     & \ref{1214b-best}         \\
Model 3    & Solar                        & Opaque clouds &              &              & 0.3        & \ref{1214b-best}         \\
Model 4    & 1\% H$_2$O, 99\% N$_2$       & None          &              &              &            & \ref{1214b-best}         \\
Model 5    & 1\% H$_2$O, 99\% N$_2$       & Tholin        & 0.01         & 10$^6$       & 0.1-0.001  & \ref{1214b-best}         \\
\enddata
\label{models}
\end{deluxetable}
\end{center}
\clearpage

\begin{figure}[htp]			
\begin{center}
\includegraphics[scale=0.9]{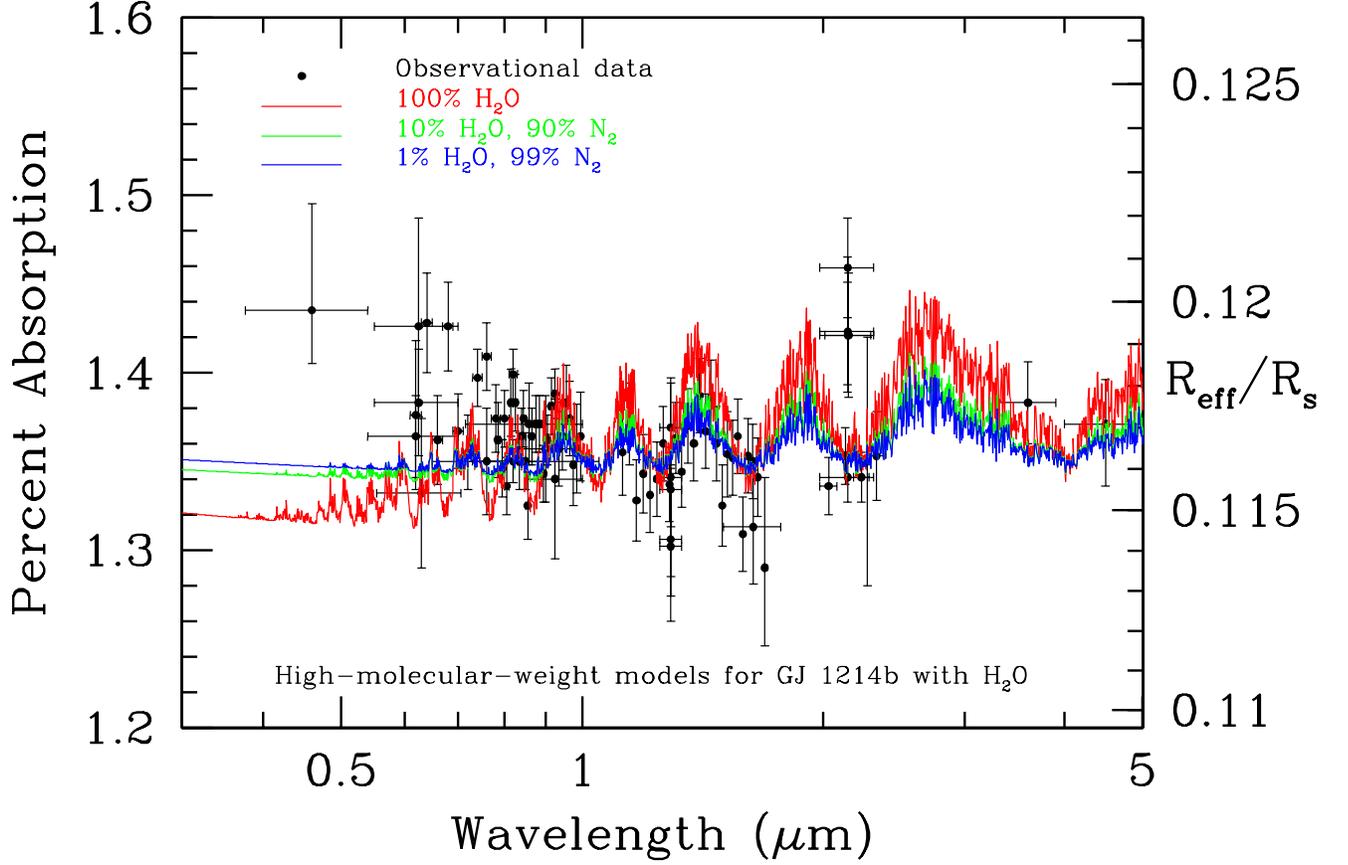} 
\end{center}
\caption{Transit depth in percent vs. wavelength in microns for atmospheric models for GJ 1214b with high molecular weights, specifically, various proportions of water with the remainder being nitrogen, which we take to have only Rayleigh scattering opacity.  With a higher molecular weight, the scale height of the atmosphere is smaller, as are the amplitudes of the spectral features.  Almost any proportion of water is consistent with most of the observed infrared features, again with the $K_s$ band being the primary exception.  The best fit covers a broad range of abundances of $\lesssim 10\%$, water, which reproduces the suspected 1.5 ${\rm \mu m}$ feature in Berta et al. (2011b) and the $K$-band measurements of Bean et al. (2011).  However, these models fail to produce the short-wavelength rise, and the lower abundances fail to produce the slight observed rise in the mid-infrared.  The molecular weights of water and nitrogen result in a scale height so small that the slope of the Rayleigh tail is much less than that of the observed short-wavelength rise.}
\label{water-model}
\end{figure}

\clearpage

\begin{thebibliography}{n}

\bibitem[Bar-Nun, Kleinfeld, \& Ganor(1988)]{Bar-Nun} Bar-Nun, A., Kleinfeld, I. \& Ganor, E. 1988, J. Geophys. Res., 93, 8383

\bibitem[Basilevsky \& Head(2003)]{Basilevsky} Basilevsky, A. T. \& Head, J. W. 2003, Rep. Prog. Phys., 66, 1699

\bibitem[Batalha et al.(2011)]{Batalha} Batalha, N. M., Borucki, W. J., Bryson, S. T. et al. 2011, ApJ, 729, 1

\bibitem[Bean, Kempton, \& Homeier(2010)]{Bean2010} Bean, J. L., Kempton, E. M.-R., \& Homeier, D. 2010, Nature, 468, 669

\bibitem[Bean et al.(2011)]{Bean2011} Bean, J. L., Desert, J.-M., Kabath, P. et al. 2011, \apj 743 92

\bibitem[Berta et al.(2011a)]{Berta} Berta, Z. K., Charbonneau, D., Bean, J., Irwin, J., et al. 2011, \apj, 736, 12

\bibitem[Berta et al.(2011b)]{Berta II} Berta, Z. K., Charbonneay, D., D\'{e}sert, J.-M., et al. 2011, accepted to \apj, (arXiv:1103.2370v1)


\bibitem[Borucki et al.(2011)]{BoruckiII} Borucki, W. J., Koch, D. G., Batalha, N., et al. 2011, \apj, 745, 120

\bibitem[Burrows \& Orton(2009)]{Burrows2009} Burrows, A. \& Orton, G. 2009, in Exoplanets, ed. S. Seager, (Space Science Series; Tuscon, AZ: Univ. Arizona Press), 419

\bibitem[Burrows \& Sharp(1999)]{Burrows} Burrows, A. \& Sharp, C. 1999, \apj, 512, 843

\bibitem[Carter et al.(2011)]{Carter} Carter, J. A., Winn, J. N., Holman, M. J., et al. 2010, \apj, 730, 82

\bibitem[Charbonneau et al.(2009)]{Charbonneau} Charbonneau, D., Berta, Z. K., Irwin, J., et al. 2009, Nature, 462, 891

\bibitem[Charbonneau et al.(2002)]{Charbonneau2002} Charbonneau, D., Brown, T. M., Noyes, R. W., \& Gilliland, R. L. 2002, \apj, 568, 377

\bibitem[Croll et al.(2011)]{Croll} Croll, B., Albert, L., Jayawardhana, R., et al. 2011, \apj, 736, 78

\bibitem[Crossfield, Barman, \& Hansen(2011)]{Crossfield} Crossfield, I. J. M., Barman, T., \& Hansen, B. M. S. 2011, \apj, 736, 132

\bibitem[De Mooij et al.(2011)]{de Mooij} De Mooij, E. J. W., Brogi, M., de Kok, R. J., et al. 2011, \aap, 538, A46

\bibitem[Deming et al.(2005)]{Deming} Deming, D., Seager, S., Richardson, L. J., \& Harrington, J. 2005, Nature, 434, 740

\bibitem[Demory et al.(2011)]{Demory} Demory, B.-O., Gillon, M., Deming, D. et al. 2011, \aap, 533, A114

\bibitem[D\'{e}sert et al.(2011)]{Desert} D\'{e}sert, J.-M., Bean, J., Kempton, E. M.-R., et al. 2011, \apj, 731, L40



\bibitem[Fortney, Marley, \& Barnes(2007)]{Fortney} Fortney, J. J., Marley, M. S., \& Barnes, J. W. 2007, \apj, 659, 1661


\bibitem[Guenther et al.(2010)]{Guenther} Guenther, E. W., Cabrera, J., Erikson, A., et al. 2010, \aap, 525, A24




\bibitem[Khare et al.(1984)]{Khare} Khare, B. N., Sagan, C., Arakawa, E. T., Suits, F., Callcott, T. A., \& Williams, M. W. 1984, Icarus 60, 127-137

\bibitem[Knutson et al.(2008)]{Knutson} Knutson, H. A., Charbonneau, D., Allen, L. E., Burrows, A., \& Megeath, S. T. 2008, \apj, 673, 526

\bibitem[Kuchner \& Seager(2005)]{Kuchner} Kuchner, M. J. \& Seager, S. 2005, \apj, submitted (arXiv:astro-ph/0504214v2)

\bibitem[Kundurthy et al.(2010)]{Kundurthy} Kundurthy, P., Agol, E., Becker, A. C., et al. 2010, \apj, 731, 123

\bibitem[Lecavelier des Etangs et al.(2008)]{Lecavelier} Lecavelier des Etangs, A., Pont, F., Vidal-Madjar, A., \& Sing, D. 2008, \aap, 481, L83

\bibitem[L\'{e}ger et al.(2009)]{Leger} L\'{e}ger, A., Rouan, D., Schieder, J., et al. 2009, \aap, 506, 287

\bibitem[Madhusudhan et al.(2010)]{Madhusudhan} Madhusudhan, N., Harrington, J., Stevenson, K. B., et al. 2010, Nature, 469, 64


\bibitem[Miller-Ricci \& Fortney(2010)]{Miller-Ricci2010} Miller-Ricci, E., \& Fortney, J. J. 2010, \apj, 716, L74

\bibitem[Miller-Ricci, Sasselov, \& Seager(2009)]{Miller-Ricci2009} Miller-Ricci, E., Sasselov, D., \& Seager, S. 2009, \apj, 690, 1056

\bibitem[Murgas et al. (2012)]{Murgas} Murgas, F., Palle, E., Cabrera-Lavers, A., et al. 2012, \aap (arXiv:1206.6619)



\bibitem[Palmer \& Williams(1975)]{Palmer} Palmer, K. F. \& Williams, D. 1975, Appl. Opt. 14, 208


\bibitem[Pont et al.(2008)]{Pont} Pont, F., Knutson, H., Gilliland, R. L., Moutou, C., \& Carbonneau, D. 2008, MNRAS, 385, 109

\bibitem[Rogers \& Seager(2010)]{Rogers} Rogers, L. A., \& Seager, S. 2010, \apj, 716, 1208

\bibitem[Sada et al.(2010)]{Sada} Sada, P. V., Deming, D., Jackson, B., et al. 2010, \apj, 720, L215

\bibitem[Sagan \& Khare(1979)]{Sagan} Sagan, C. \& Khare, B. N. 1979, Nature, 277, 102


\bibitem[Seager et al.(2007)]{Seager} Seager, S., Kuchner, M., Hier-Majumder, C., \& Militzer, B. 2007, \apj, 669, 1279

\bibitem[Sharp \& Burrows(2007)]{Sharp} Sharp, C. M., \& Burrows, A. 2007, \apjs, 168, 140


\bibitem[Sotin, Grasset, \& Mocquet(2007)]{Sotin} Sotin, C., Grasset, O., \& Mocquet, A. 2007, Icarus, 191, 337

\bibitem[Sudarsky, Burrows, \& Hubeny(2003)]{Sudarsky} Sudarsky, D., Burrows, A., \& Hubeny, I. 2003, \apj, 588, 1121

\bibitem[Swain, Vasisht, \& Tinetti(2008)]{Swain} Swain, M. R., Vasisht, G., \& Tinetti, G. 2008, Nature 452, 329

\bibitem[Valencia, Sasselov, \& O'Connell(2007)]{Valencia} Valencia, D., Sasselov, D. D., \& O'Connell, R. J. 2007, \apj 665, 1413

\bibitem[Von Braun et al.(2011)]{Von Braun} Von Braun, K., Boyajian, T. S., ten Brummelaar, T. A., et al. 2011, \apj, 740, 49

\bibitem[Winn et al.(2011)]{Winn} Winn, J. N., Matthews, J. M., Dawson, R. I., et al. 2011, \apj, 737, L18

\end{thebibliography}
\end{document}